\documentclass[apj, numberedappendix]{emulateapj}
\usepackage{xspace,multirow}

\usepackage{hyperref}
\usepackage{color}

\newcommand{\ang}{\AA\xspace}

\usepackage{tikz}
\usetikzlibrary{shapes.geometric, arrows}

\slugcomment{Accepted for Publication in the Astronomical Journal.}
\shorttitle{NICMOS 2 Calibration}
\shortauthors{Rubin et al.}

\begin{document}

\title{A Calibration of NICMOS Camera 2 for Low Count-Rates\footnotemark[1]}

\footnotetext[1]{Based on observations with the NASA/ESA \textit{Hubble Space Telescope}, obtained at the Space Telescope Science Institute, which is operated by AURA, Inc., under NASA contract NAS 5-26555, under programs SM2/NIC-7049, SM2/NIC-7152, CAL/NIC-7607, CAL/NIC-7691, CAL/NIC-7693, GO-7887, CAL/NIC-7902, CAL/NIC-7904, GO/DD-7941, SM3/NIC-8983, SM3/NIC-8986, GTO/ACS-9290, ENG/NIC-9324, CAL/NIC-9325, GO-9352, GO-9375, SNAP-9485, CAL/NIC-9639, GO-9717, GO-9834, GO-9856, CAL/NIC-9995, CAL/NIC-9997, GO-10189, GO-10258, CAL/NIC-10381, CAL/NIC-10454, GO-10496, CAL/NIC-10725, CAL/NIC-10726, GO-10886, CAL/NIC-11060, CAL/NIC-11061, GO-11135, GO-11143, GO-11202, CAL/NIC-11319, GO/DD-11359, SM4/WFC3-11439, SM4/WFC3-11451, GO-11557, GO-11591, GO-11600, GO/DD-11799, CAL/WFC3-11921, CAL/WFC3-11926, GO/DD-12051, GO-12061, GO-12062, GO-12177, CAL/WFC3-12333, CAL/WFC3-12334, CAL/WFC3-12341, GO-12443, GO-12444, GO-12445, CAL/WFC3-12698, CAL/WFC3-12699, GO-12874, CAL/WFC3-13088, and CAL/WFC3-13089.}

\author{
D. Rubin\altaffilmark{2},
G. Aldering\altaffilmark{3},
R. Amanullah\altaffilmark{4},
K. Barbary\altaffilmark{3},
K. S. Dawson\altaffilmark{5},
S. Deustua\altaffilmark{6},
L. Faccioli\altaffilmark{7, 8},
V. Fadeyev\altaffilmark{9},
H. K. Fakhouri\altaffilmark{3, 10},
A. S. Fruchter\altaffilmark{6},
M. D. Gladders\altaffilmark{11, 12},
R. S. de Jong\altaffilmark{13},
A. Koekemoer\altaffilmark{6},
E. Krechmer\altaffilmark{10},
C. Lidman\altaffilmark{14},
J. Meyers\altaffilmark{15},
J. Nordin\altaffilmark{3, 16},
S. Perlmutter\altaffilmark{3, 10},
P. Ripoche\altaffilmark{17},
D. J. Schlegel\altaffilmark{3},
A. Spadafora\altaffilmark{3},
N. Suzuki\altaffilmark{18}
\\(The Supernova Cosmology Project)}

\altaffiltext{2}{Department of Physics, Florida State University, Tallahassee, FL, 32306}
\altaffiltext{3}{E.O. Lawrence Berkeley National Lab, 1 Cyclotron Rd., Berkeley, CA, 94720}
\altaffiltext{4}{The Oskar Klein Centre, Department of Physics, AlbaNova, Stockholm University, SE-106 91 Stockholm, Sweden}
\altaffiltext{5}{Department of Physics and Astronomy, University of Utah, 115 S 1400 E, Salt Lake City, UT 84112}
\altaffiltext{6}{Space Telescope Science Institute, 3700 San Martin Drive, Baltimore, MD 21218}
\altaffiltext{7}{Kavli Institute for Astronomy and Astrophysics, Peking University, Beijing 100871, P. R. China}
\altaffiltext{8}{National Astronomical Observatories, Chinese Academy of Sciences, Beijing 100012, P. R. China}
\altaffiltext{9}{Santa Cruz Institute for Particle Physics, University of California, Santa Cruz, 1156 High Street, Santa Cruz, CA 95064}
\altaffiltext{10}{Department of Physics, University of California Berkeley, Berkeley, CA 94720}
\altaffiltext{11}{Department of Astronomy and Astrophysics, University of Chicago, 5640 S Ellis Ave, Chicago, IL 60637}
\altaffiltext{12}{Kavli Institute for Cosmological Physics, The University of Chicago, 5640 South Ellis Avenue, Chicago, IL 60637}
\altaffiltext{13}{Leibniz-Institut f\"ur Astrophysik Potsdam (AIP), An der Sternwarte 16, D-14482, Potsdam, Germany}
\altaffiltext{14}{Australian Astronomical Observatory, PO Box 296, Epping, NSW 1710, Australia}
\altaffiltext{15}{Department of Physics, Stanford University, 450 Serra Mall, Stanford, CA 94305}
\altaffiltext{16}{Space Sciences Lab, 7 Gauss Way, Berkeley, CA 94720}
\altaffiltext{17}{Google, Pittsburgh, PA}
\altaffiltext{18}{Kavli Institute for the Physics and Mathematics of the Universe, University of Tokyo, Kashiwa, 277-8583, Japan}

\newcommand{\JWeight}{1~mmag\xspace}

\newcommand{\HWeight}{$<1$~mmag\xspace}

\newcommand{\ErrorHStat}{6~mmag\xspace}

\newcommand{\Hfixalphadiff}{3~mmag\xspace}

\newcommand{\ErrorHPSF}{7~mmag\xspace}

\newcommand{\ErrorHCent}{$<1$~mmag\xspace}

\newcommand{\ErrorHAnn}{1~mmag\xspace}

\newcommand{\ErrorHTem}{12~mmag\xspace}

\newcommand{\ErrorHEBV}{2~mmag\xspace}
\newcommand{\ErrorHEE}{2~mmag\xspace}
\newcommand{\ErrorHColCol}{2~mmag\xspace}
\newcommand{\RMSHColCol}{10~mmag\xspace}
\newcommand{\ErrorHTemTotal}{13~mmag\xspace}
\newcommand{\ErrorHTotal}{17~mmag\xspace}

\newcommand{\ErrorJStat}{10~mmag\xspace}

\newcommand{\Jfixalphadiff}{4~mmag\xspace}

\newcommand{\Jfitbetadiff}{10~mmag\xspace}
\newcommand{\Jfitbetabeta}{$8.7\pm6.2$}

\newcommand{\ErrorJPSF}{8~mmag\xspace}

\newcommand{\ErrorJCent}{1~mmag\xspace}

\newcommand{\ErrorJAnn}{5~mmag\xspace}

\newcommand{\ErrorJTem}{$<1$~mmag\xspace}

\newcommand{\ErrorJEBV}{$<1$~mmag\xspace}
\newcommand{\ErrorJEE}{2~mmag\xspace}
\newcommand{\ErrorJColCol}{1~mmag\xspace}
\newcommand{\RMSJColCol}{5~mmag\xspace}
\newcommand{\ErrorJTemTotal}{3~mmag\xspace}
\newcommand{\ErrorJTotal}{14~mmag\xspace}

\newcommand{\ErrorHpreStat}{8~mmag\xspace}

\newcommand{\ErrorHprePSF}{9~mmag\xspace}

\newcommand{\ErrorHpreCent}{1~mmag\xspace}

\newcommand{\ErrorHpreAnn}{1~mmag\xspace}

\newcommand{\ErrorHpreTem}{4~mmag\xspace}

\newcommand{\ErrorHpreEBV}{1~mmag\xspace}
\newcommand{\ErrorHpreEE}{2~mmag\xspace}
\newcommand{\ErrorHpreColCol}{2~mmag\xspace}
\newcommand{\RMSHpreColCol}{10~mmag\xspace}
\newcommand{\ErrorHpreTemTotal}{7~mmag\xspace}
\newcommand{\ErrorHpreTotal}{14~mmag\xspace}

\newcommand{\ErrorJpreStat}{8~mmag\xspace}

\newcommand{\ErrorJprePSF}{13~mmag\xspace}

\newcommand{\ErrorJpreCent}{$<1$~mmag\xspace}

\newcommand{\ErrorJpreAnn}{10~mmag\xspace}

\newcommand{\ErrorJpreTem}{4~mmag\xspace}

\newcommand{\ErrorJpreEBV}{3~mmag\xspace}
\newcommand{\ErrorJpreEE}{2~mmag\xspace}
\newcommand{\ErrorJpreColCol}{1~mmag\xspace}
\newcommand{\RMSJpreColCol}{5~mmag\xspace}
\newcommand{\ErrorJpreTemTotal}{6~mmag\xspace}
\newcommand{\ErrorJpreTotal}{19~mmag\xspace}

\newcommand{\HZPOffsetrev}{$-2.378$~mag\xspace}
\newcommand{\NICHSTZPrev}{$25.803$\xspace}

\newcommand{\JZPOffsetrev}{$-3.138$~mag\xspace}
\newcommand{\NICJSTZPrev}{$25.296$\xspace}

\newcommand{\HpreZPOffsetrev}{$-2.683$~mag\xspace}
\newcommand{\NICHpreSTZPrev}{$25.498$\xspace}

\newcommand{\JpreZPOffsetrev}{$-3.591$~mag\xspace}
\newcommand{\NICJpreSTZPrev}{$24.843$\xspace}

\newcommand{\HZPOffsetsyn}{$-2.376$~mag\xspace}
\newcommand{\NICHSTZPsyn}{$25.789$\xspace}

\newcommand{\JZPOffsetsyn}{$-3.145$~mag\xspace}
\newcommand{\NICJSTZPsyn}{$25.272$\xspace}

\newcommand{\HpreZPOffsetsyn}{$-2.679$~mag\xspace}
\newcommand{\NICHpreSTZPsyn}{$25.487$\xspace}

\newcommand{\JpreZPOffsetsyn}{$-3.592$~mag\xspace}
\newcommand{\NICJpreSTZPsyn}{$24.825$\xspace}

\newcommand{\StarsJ}{$-3.16\pm0.04$\xspace}

\newcommand{\given}{standard\xspace}
\newcommand{\suggested}{revised\xspace}
\newcommand{\Given}{Standard\xspace}
\newcommand{\Suggested}{Revised\xspace}

\newcommand{\spatialCRL}{$\sim 10$ percent\xspace} 
\newcommand{\WFCWFCscatter}{0.03\xspace}

\newcommand{\PascalZPOffset}{0.055\xspace}

\newcommand{\TotalHGalaxies}{28\xspace}
\newcommand{\TotalHGalaxieswithSpec}{14\xspace}

\newcommand{\ErrorJWave}{30\ang\xspace}
\newcommand{\ErrorHWave}{17\ang\xspace} 

\newcommand{\averageHflux}{0.16 ADU/s\xpace}
\newcommand{\averageJflux}{0.35 ADU/s\xpace} 

\newcommand{\meanJnlc}{0.23~magnitudes\xspace}
\newcommand{\meanHnlc}{0.10~magnitudes\xspace}

\newcommand{\stsciJSTZPpostNCS}{25.262\xspace} 
\newcommand{\stsciJSTZPpreNCS}{24.815\xspace} 
\newcommand{\stsciHSTZPpostNCS}{25.799\xspace} 
\newcommand{\stsciHSTZPpreNCS}{25.470\xspace} 

\newcommand{\NICJEE}{0.935\xspace}
\newcommand{\NICHEE}{0.917\xspace}

\begin{abstract}

NICMOS 2 observations are crucial for constraining distances to most of the existing sample of $z > 1$ SNe~Ia. Unlike the conventional calibration programs, these observations involve long exposure times and low count rates.
Reciprocity failure is known to exist in HgCdTe devices and a correction for this effect has already been implemented for high and medium count-rates. However observations at faint count-rates rely on extrapolations. Here instead, we provide a new zeropoint calibration directly applicable to faint sources.
This is obtained via inter-calibration of NIC2 F110W/F160W with WFC3 in the low count-rate regime using $z \sim 1$ elliptical galaxies as tertiary calibrators. These objects have relatively simple near-IR SEDs, uniform colors, and their extended nature gives superior signal-to-noise at the same count rate than would stars. The use of extended objects also allows greater tolerances on PSF profiles. We find ST magnitude zeropoints (after the installation of the NICMOS cooling system, NCS) of $\NICJSTZPrev \pm 0.022$ for F110W and $\NICHSTZPrev \pm 0.023$ for F160W, both in agreement with the calibration extrapolated from count-rates $\gtrsim$ 1,000 times larger (\stsciJSTZPpostNCS and \stsciHSTZPpostNCS). Before the installation of the NCS, we find $\NICJpreSTZPrev \pm 0.025$ for F110W and $\NICHpreSTZPrev \pm 0.021$ for F160W, also in agreement with the high-count-rate calibration (\stsciJSTZPpreNCS and \stsciHSTZPpreNCS). We also check the \given bandpasses of WFC3 and NICMOS 2 using a range of stars and galaxies at different colors and find mild tension for WFC3, limiting the accuracy of the zeropoints. To avoid human bias, our cross-calibration was ``blinded'' in that the fitted zeropoint differences were hidden until the analysis was finalized.

\end{abstract}


\section{Introduction}

With the installation in 1997 of the Near Infrared Camera and Multi-Object Spectrometer (NICMOS) instrument, the Hubble Space Telescope (HST) first gained powerful near-IR capabilities \citep{thompson92, viana09}. With low sky and diffraction-limited imaging, NICMOS was $\sim 10$ times faster at $J$ and $H$ point-source imaging than large ground-based telescopes with adaptive optics. Three cameras were available (NIC1, NIC2, and NIC3), each 256$\times$256 pixels, with pixel sizes of 0\farcs043 (NIC1), 0\farcs075 (NIC2), and 0\farcs2 (NIC3, which also had grism spectroscopy). The instrument was originally cooled to 61K by a block of nitrogen ice until lack of coolant stopped operations in 1999. In 2002, a servicing mission installed a cryocooler (the NICMOS Cooling System, NCS), allowing consumable-free operations at 77K.

NICMOS enabled the first probes of the earliest half of the expansion history of the universe \citep{riess01, riess04, riess07, suzuki12, rubin13}. Although precision ground-based $z>1$ SN measurements are possible \citep[][Rubin et al., in prep]{tonry03, amanullah10, suzuki12}, the required long exposure times with 10m-class telescopes make building a large sample expensive. NICMOS allowed for the measurement of precision colors (and thus, distances) for these distant SNe, sampling the rest-frame $B, V, R,$ or $I$ band, depending on filter and redshift. Even with the forthcoming Wide Field Camera 3 (WFC3)-observed SNe \citep{graur14, rodney14} \citetext{\citealp[from the Cluster Lensing And Supernova search with Hubble, CLASH:][]{postman12}, and \citealp[the Cosmic Assembly Near-infrared Deep Extragalactic Legacy Survey, CANDELS:][]{grogin11, koekemoer11}}, NICMOS-observed SNe~Ia will continue to make up the bulk of the $z>1$ sample.

NICMOS has proven to be a challenging instrument to calibrate. \citet{bohlin0502} first found evidence of a count-rate non-linearity (CRNL) in NIC3 when extending spectrophotometric standards into the near IR. The Space Telescope Imaging Spectrograph (STIS) and NICMOS showed clear disagreement over the wavelength range in common (8,000 to 10,000\ang), with NIC3 indicating a relative deficit of flux for fainter sources. Parameterizing the CRNL in terms of relative magnitude deficit per dex (factor of ten in count rate), NIC3 showed an increase of 0.06 mag/dex for count rates from $\sim 2$ to $\sim 3,000$/s ($\sim 0.18$ magnitudes over this $\sim 3$ dex range). Spectroscopy and imaging from the HST Advanced Camera for Surveys (ACS) agreed with STIS, pointing to NIC3 as the root of the problem. A comparison of three white dwarfs against models showed a strong wavelength dependence to the CRNL, with the CRNL consistent with zero longward of 16,000\ang.

\citet{mobasher0503} first investigated this effect with ground-based data using both stars and galaxies. The stars had been observed in both F110W (a broad filter spanning $Y$ and $J$ centered at $1.1\mu$m) and F160W (similar to $H$, centered at $1.6\mu$m) in NIC2 and with ground-based $J$ and $H$ over the $J$ magnitude range 8-17 \citep{stephens00}. Their galaxies ranged in brightness down to the sky level, and were likewise observed in $J$ and $H$, but the NICMOS data came from NIC3 instead of NIC2. The star measurements showed no significant CRNL in either NIC2 band, with the F110W CRNL constrained to be a factor of at least 2-3 smaller than the NIC3 result from \citet{bohlin0502}. The galaxy measurements showed no significant CRNL until the measurements approached the sky level ($J \sim 23$) when the scatter became large and offsets $\sim 0.1$ magnitudes may have been indicated. The authors suggest that charge trapping may be responsible for the observed CRNL: exposures $\gtrsim 155$s (the persistence timescale), used to measure faint objects, may be able to fully fill the traps, resulting in a smaller CRNL.

\citet{dejong0601} used exposures of star fields with and without counts enhanced by a flatfield lamp to directly measure the linearity of NIC1 and NIC2. Only count rates between $\sim 50$ and 2,000 counts per second were probed by this technique in NIC2 F110W, but the CRNL again seemed to be roughly constant in mag/dex over this range. Interestingly, the NIC2 F110W CRNL seemed to be the same size before and after the installation of the NCS and the associated change in temperature. In conflict with the exposure-time/charge-trap hypothesis, the observed CRNL is the same size whether the lamp-off data are taken after the lamp-on data (when the charge traps should be full) or before. In addition, \citet{bohlin0602} checked the \citet{bohlin0502} analysis using longer grism exposure times. They also find the same size NIC3 CRNL as with the shorter exposures, again at odds with the \citet{mobasher0503} results and a simple picture of charge-trapping. \citep[We do however note that more-detailed models of charge-trapping do seem to fit lab-measured data, see][]{regan12}.

Taking the measurements from \citet{dejong0601} and \citet{bohlin0602}, \citet{dejong0603} introduced a routine, \texttt{rnlincor}, that corrects the values in an image using an assumed power-law relation between the corrected and original values. The power-law is parameterized in units of mag/dex in the sense that
\begin{eqnarray*}
\mathrm{CR}_{\mathrm{estimated}} & = &  \mathrm{CR}^{1/\{0.063 \mathrm{[mag/dex}]/2.5 + 1\}}_{\mathrm{observed}} \hspace{0.2in} \mathrm{for\ F110W}\\
\mathrm{CR}_{\mathrm{estimated}} & = & \mathrm{CR}^{1/\{0.029 \mathrm{[mag/dex}]/2.5 + 1\}}_{\mathrm{observed}} \hspace{0.2in} \mathrm{for\ F160W.}
\end{eqnarray*}

The current convention is to then use the corrected count rate in combination with the zeropoint provided from bright standard stars. This procedure was used to calibrate the SNe in the Great Observatories Origins Deep Survey (GOODS) fields \citep{riess07} and the Supernova Cosmology Project (SCP) high-redshift SNe in \citet{nobili09}.

However, this solution was not an adequate calibration. The 0.006 mag/dex uncertainty on the NIC2 F110W CRNL translates into a $\sim 0.024$ mag uncertainty over the 4 dex range between the standard stars and high-redshift SNe. The effect of the strong wavelength-dependence of the CRNL over the F110W filter is hard to model for faint sources, as the amplifier glow is not at the same effective wavelength as the observations, and the dark current has no wavelength. As these sources are a significant fraction of the total background, this introduces $\sim 0.02$ magnitudes of uncertainty. It is also unclear even what effective count rate faint observations are taken at, as the amplifier glow may be constant, or produced in short bursts of high counts/second. We note that the \citet{mobasher0503} results could indicate that the NIC3 F110W power law breaks down at low count-rates and is wrong by $\sim 0.1$ magnitudes at low count-rates (possibly the sum of the above effects).

Given these issues, we were awarded 14 orbits\footnote{GO/DD-11799 and GO/DD-12051} to complete a precision calibration of NIC2 F110W at low count-rates, unlocking the full potential of the high-redshift SN Ia data. \citet{suzuki12} and \citet{rubin13} relied on a first-round SCP F110W calibration against a combination of ACS WFC and deep ground-based $J$ and $K$ data. This calibration indicated a zeropoint \PascalZPOffset magnitudes fainter (larger) than the extrapolation of the higher-count-rate calibrations, showing a weakening of the CRNL at low count-rates. Here, we derive an updated result, taking advantage of the similar WFC3 IR bandpasses. Given the larger number of archival WFC3 and NICMOS F160W observations now available, we are also able derive a result for F160W. Using fortuitous archival observations of mid-redshift galaxy clusters, we make the same measurement for pre-NCS observations.

Concurrently with our NIC2 investigations, \citet{riess10a} compared WFC3 IR starfield data against ACS F850LP and NIC2 F110W and F160W. With good precision (but only for count rates that are more than 10 times higher than high-redshift SN count rates) WFC3 IR showed a small power-law index CRNL ($\sim 6$ times smaller than for NIC2 F110W), that is approximately constant in mag/dex (as a function of count-rate) when compared against ACS and \texttt{rnlincor}-corrected NIC2 images. Similar WFC3 IR CRNL measurements were made by \citet{riess10b, riess11} independently of NICMOS, so \texttt{rnlincor} seems to be accurate within the given uncertainties at these count rates.

\begin{figure}[ht]
\begin{center}
\includegraphics[width=3.5in]{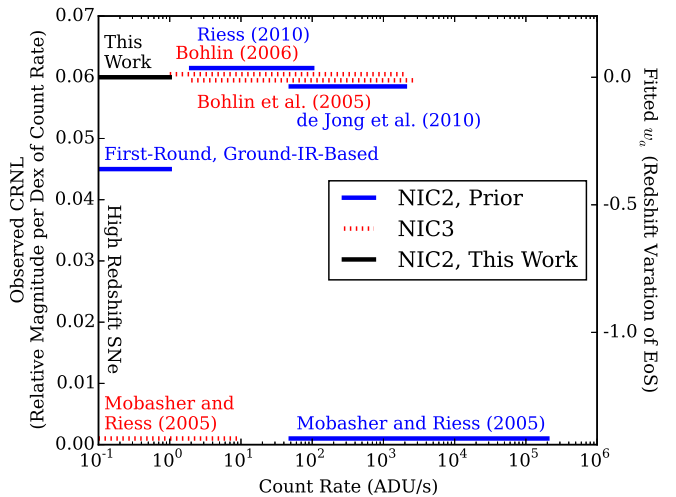}
\end{center}
\caption{Visual summary of the referenced NICMOS F110W calibration results and their approximate cosmological implications. Although this work is concerned with NIC2, we include NIC3 results (red dotted lines) to establish the level of uncertainty in the behavior of the count-rate non-linearity (although these results are not taken at quite the same effective wavelengths as the NIC2 results). Previous NIC2 results are color-coded in blue. Each line indicates the measured CRNL index and the range of count-rates it was measured at. The ``first-round'' result indicated a fainter NIC2 F110W zeropoint at low count-rates than the other calibrations, which we plot here assuming that the CRNL has a constant size for all count-rates. The results of our new calibration are consistent with the results of \citet{dejong0601, riess10a} and are plotted in black. The cosmological results shown on the right axis are evaluated by fitting a time-varying $w_0$-$w_a$ model to the Union2.1 supernova compilation \citep{suzuki12}, and aligning $w_a=0$ with our calibration.}\label{fig:summary}
\end{figure}

Figure \ref{fig:summary} summarizes the measurements we reference. The size of the CRNL is shown (left axis), plotted against the range of count-rates over which it was measured. On the right axis, we use the Union2.1 supernova compilation, combined with BAO, CMB, and $H_0$ measurements \citep[described in more detail in][]{suzuki12} to convert from CRNL size to cosmological impact. For evaluating this impact, we use the $w_0$-$w_a$ model \citep{chevallier01, linder03} in which the equation of state parameter of dark energy smoothly varies with time as $w(a) = w_0 + w_a (1 - a)$. The high-redshift supernovae are particularly useful in constraining the time-variation, so we judge the impact using shifts in the best-fit $w_a$. A full cosmological analysis will be presented with other improvements in a future paper; for now, we compute the linear response of $w_a$ to the calibration and display that linear scale. The range of calibrations referenced here span $\sim 1.5$ in $w_a$. This is twice the size of all the other statistical and systematic uncertainties combined.

As the \texttt{rnlincor} power-law count-rate correction seems to be accurate at high count-rates, our strategy was to begin by correcting the NICMOS data for this relation. As all of our data (described below) encompasses a relatively narrow range in count-rates (centered around the count-rates of high-redshift SNe), we choose to derive an effective set of zeropoint differences between NIC2 and WFC3 (four, for F110W/F160W and pre-NCS/NCS). This strategy captures the relevant low-count-rate calibration, without necessitating the interpretation of data in other count-rate and exposure-time regimes.

\begin{figure}[h]
\begin{center}
\includegraphics[width=3.5in]{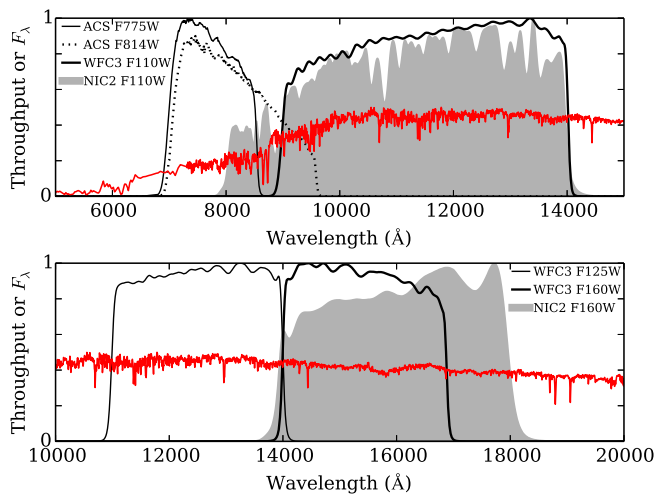}
\end{center}
\caption{The filter bandpasses referenced in this analysis, plotted against wavelength. Left to right in the top panel are the ACS WFC F775W filter (thin solid line), ACS WFC F814W filter (dotted line), the NIC2 F110W filter (filled), and the WFC3 F110W filter (thick solid line). Left to right in the bottom panel are the WFC3 F125W filter (thin solid line), the WFC3 F160W filter (thick solid line), and the NIC2 F160W filter (filled). For reference, an elliptical galaxy template redshifted to $z=1.2$ is overplotted in red. All normalizations are arbitrary.}\label{fig:bandpasses}
\end{figure}

\begin{figure}[h]
\begin{center}
\includegraphics[width=0.5 \textwidth]{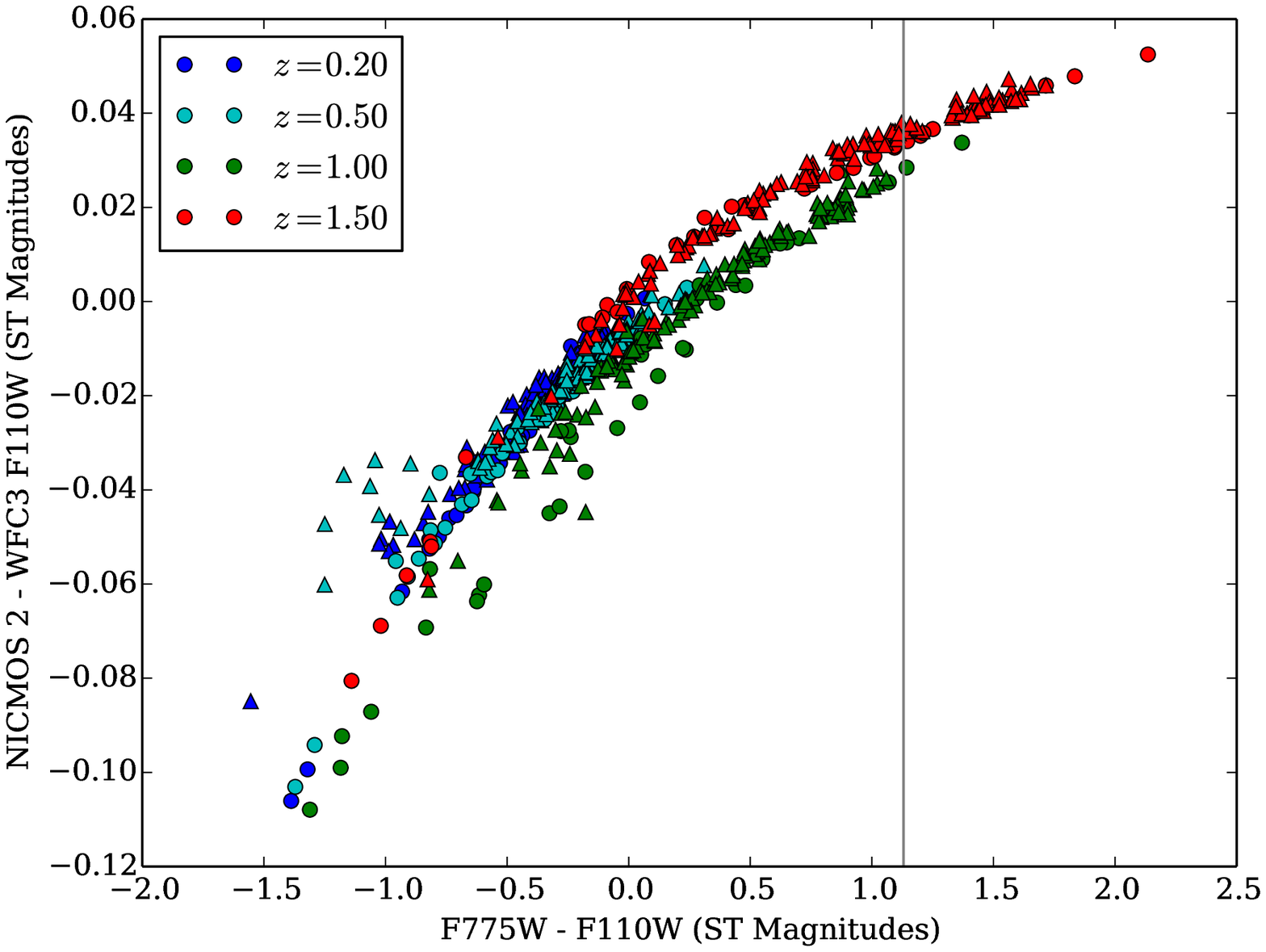}\\
\includegraphics[width=0.5 \textwidth]{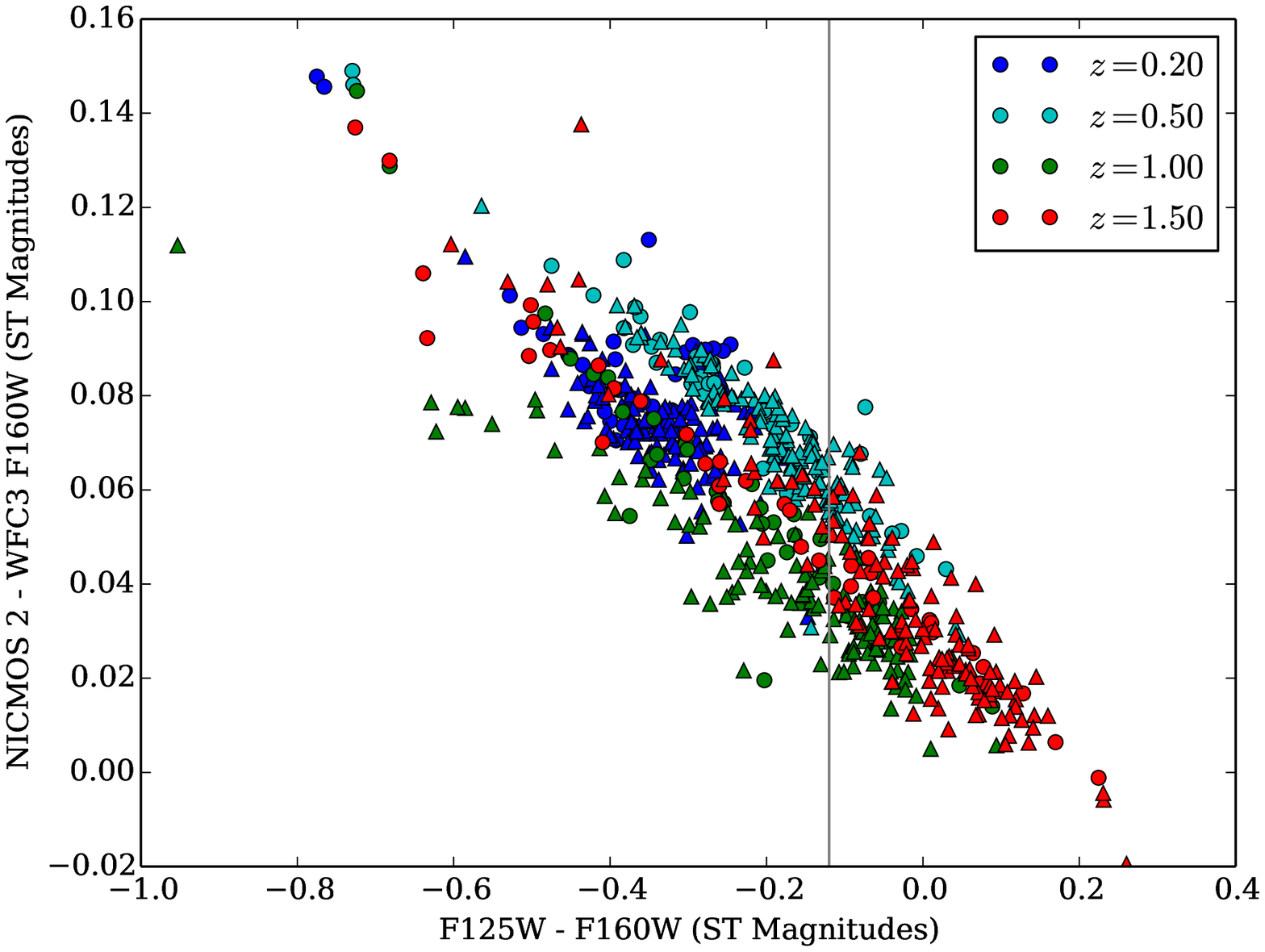}
\end{center}
\caption{Galaxy color-color relations for cross-calibrating F110W (top panel) and F160W (bottom panel), shown at a range of redshifts. Circular markers are stellar population synthesis models from \citet{bruzual03}, while triangular markers are measurements of nearby galaxies (of all types) from \citet{brown14}. The vertical gray lines indicate the median colors of the galaxies used in our calibration (for which these colors are available).}

\label{fig:colorcolor}
\end{figure}

\section{Data}

Figure~\ref{fig:bandpasses} presents the unnormalized HST bandpasses referenced in this analysis. The NIC2 and WFC3 F110W bandpasses are quite similar. The NIC2 F160W bandpass extends redder than the WFC3 F160W bandpass and requires a mild extrapolation outside the wavelength range of WFC3. In both cases, there is  enough overlap that simple color-color relations can be used to cross-calibrate NICMOS and WFC3. For the F110W calibration, we use the F775W$-$F110W color as the abscissa, except when F775W is not available and we use F814W$-$F110W. For simplicity, we avoid F850LP data, as CCD scattering makes the PSF quite color-dependent \citep{sirianni98}. For the F160W calibration, we use the WFC3 F125W$-$F160W color as the abscissa, unless F125W is not available, in which case we use F814W$-$F160W or F110W$-$F160W. Example galaxy color-color relations at a range of redshifts are shown in Figure~\ref{fig:colorcolor}.

As illustrated in Figure~\ref{fig:bandpasses}, the elliptical galaxy templates used in this analysis are relatively flat in $f_{\lambda}$ inside the F110W and F160W bandpasses. We therefore choose to conduct our cross-calibration using Space Telescope (ST) magnitudes \citep[magnitudes that are flat in $f_{\lambda}$, see][]{koorneef86}. Selecting a different magnitude system (e.g., AB or Vega, both of which use bluer references than ST) would have resulted in different calibration offsets and different correlations between calibration offsets and bandpass uncertainties. However, any cosmological results (using those cross-calibrations and covariance matrices) would be the same. ST magnitudes have the convenient advantage that the correlations can essentially be neglected.

\begin{figure*}[htp]
\includegraphics[width =  1 \textwidth]{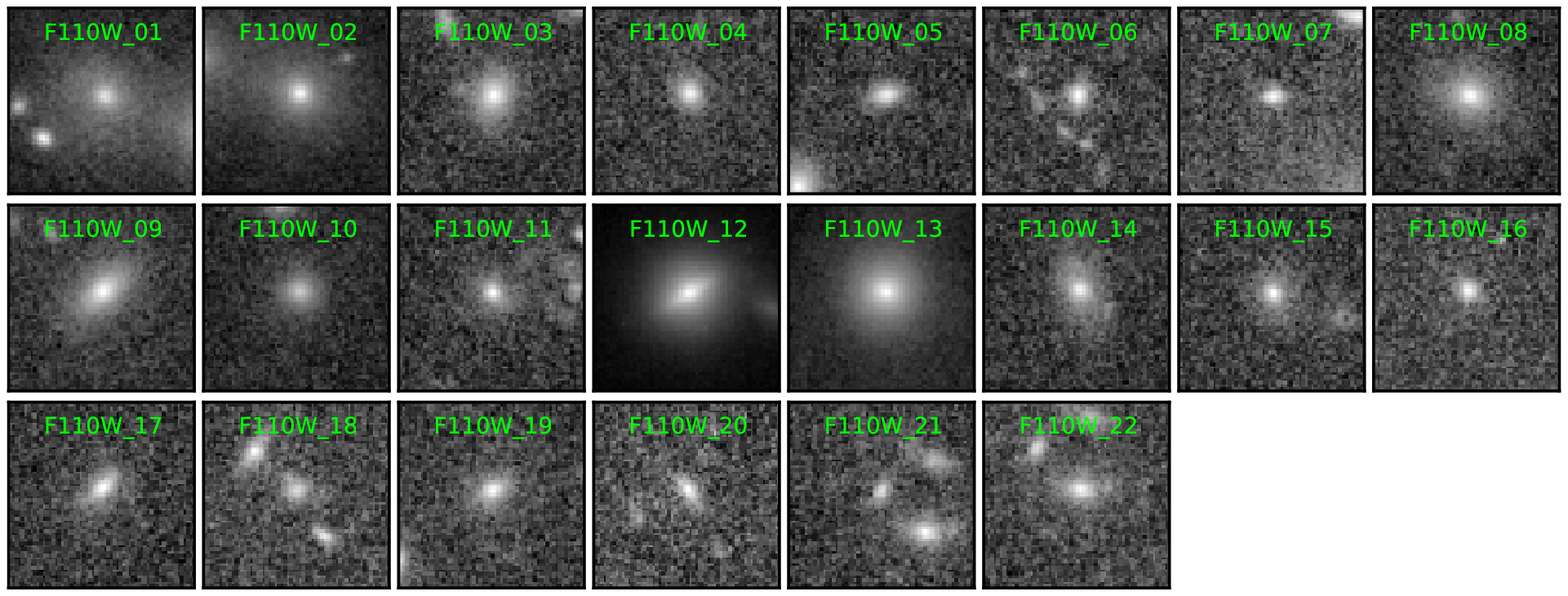}

\includegraphics[width =  1 \textwidth]{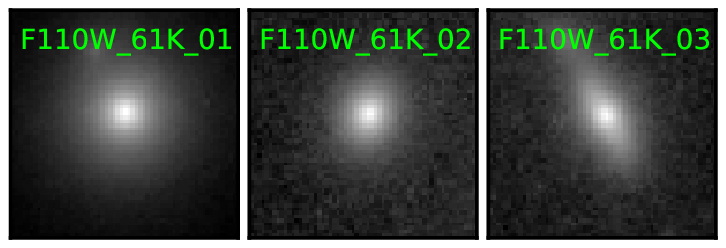}

\includegraphics[width = 1 \textwidth]{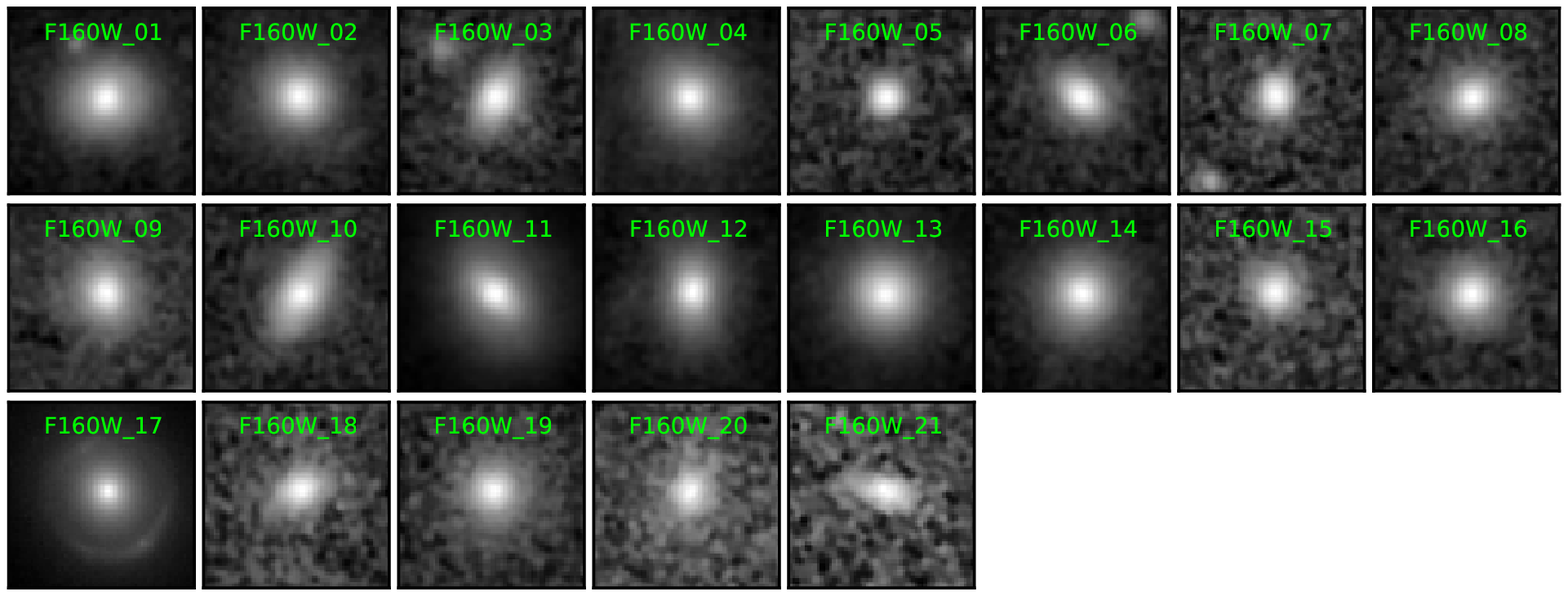}

\includegraphics[width = 1 \textwidth]{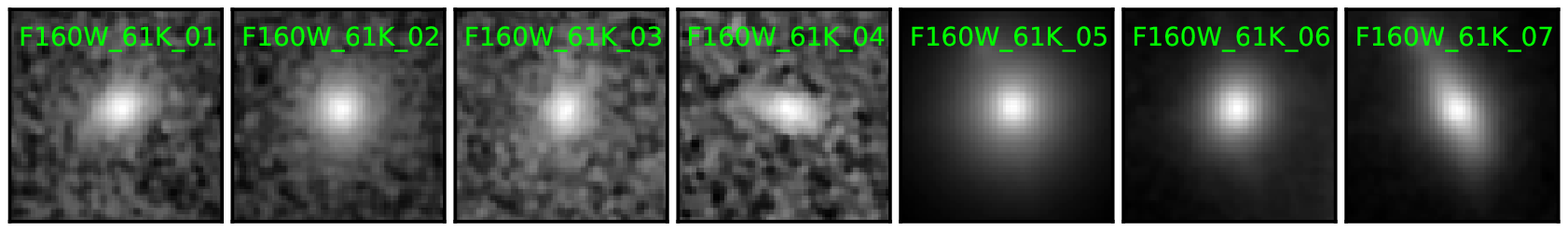}
\caption{3\arcsec\xspace by 3\arcsec\xspace cutouts around each galaxy. The scaling is sinh$^{-1}$, so it approaches $\pm$logarithmic at large absolute fluxes, while approaching linear at small fluxes. This non-linear scaling brings out faint features, such as the Einstein ring around F160W\_17.}
\label{fig:postagestamps}
\end{figure*}

We selected our calibration galaxies from ACS images, looking by eye for early-type morphologies and uniform colors. Each of the galaxies we selected showed stable colors when using photometry with different radius ranges (see Section~\ref{sec:instrumentalcolor}). Stacking the ACS data for each galaxy and removing an azimuthally symmetric galaxy model (one allowed to have ellipticity and an arbitrary spline radial profile) revealed spiral structure in some galaxies; these galaxies were removed from this analysis. For \TotalHGalaxieswithSpec out of \TotalHGalaxies galaxies in the F160W calibration, we found archival WFC3 G141 spectroscopy (covering 11000\ang to 17000\ang), allowing us to examine the near-IR SED and determine the redshift. (Many redshifts also came from the literature, as summarized in Table~\ref{tab:obs}.) For F110W, where the scatter of the color-color relation is smaller (and thus robust if a redshift is incorrect\footnote{In fact, simply assuming all F110W galaxies in the post-NCS calibration are at redshift 1.2 only changes the derived calibration by 1~milli-magnitude (mmag, 0.001 magnitudes).}), we selected red-sequence galaxies \citep[presented for the $z>1$ clusters in][]{meyers12} from the galaxy clusters ISCS J1434.4+3426 \citep{brodwin06}, RDCS J1252.9-2927 \citep{rosati04}, XMMU J2235.3-2557 \citep{mullis05}, and Abell 1835 \citep{abell89}. Images of the selected galaxies are shown in Figure~\ref{fig:postagestamps}. Among our calibration galaxies are the host galaxies for the SNe SCP06C0, SCP06H5 \citep{suzuki12}, 05Lan, 04Tha, and 05Red \citep{riess07}. For the supernovae blended with their host galaxies, we used only the supernova-free reference images in this analysis.

\clearpage

\LongTables
\begin{deluxetable*}{lrr l  r p{3.2cm}  p{1 cm} r}
\tablecolumns{8}

\tabletypesize{\footnotesize}
\setlength{\tabcolsep}{0.02in}
\tablecaption{Galaxies used in this measurement\label{tab:obs}}
\tablehead{
 \colhead{Galaxy} &
 \colhead{RA} &
 \colhead{DEC} &
 \colhead{PIDs} &
 \colhead{Redshift} &
 \colhead{Redshift Source} &
 \colhead{Emission$^1$} &
 \colhead{MW E($B-V$)} }

\startdata

\cutinhead{F110W, Post-NCS}
F110W\_01 & 193.22757 & $ -29.45461 $ & s, r, c & 1.24 & RDCS J1252.9$-$2927 & \nodata & 0.075 \\
F110W\_02 & 193.22703 & $ -29.45479 $ & s, r, c & 1.24 & RDCS J1252.9$-$2927 & \nodata & 0.075 \\
F110W\_03 & 193.22706 & $ -29.45644 $ & s, r, c & 1.24 & RDCS J1252.9$-$2927 & \nodata & 0.075 \\
F110W\_04 & 193.23039 & $ -29.45358 $ & s, r, c & 1.24 & RDCS J1252.9$-$2927 & \nodata & 0.075 \\
F110W\_05 & 193.22661 & $ -29.45602 $ & s, r, c & 1.24 & RDCS J1252.9$-$2927 & \nodata & 0.075 \\
F110W\_06 & 193.22575 & $ -29.45325 $ & s, r, c & 1.24 & RDCS J1252.9$-$2927 & \nodata & 0.075 \\
F110W\_07 & 193.22816 & $ -29.45401 $ & s, r, c & 1.24 & RDCS J1252.9$-$2927 & \nodata & 0.075 \\
F110W\_08 & 193.23039 & $ -29.45451 $ & s, r, c & 1.24 & RDCS J1252.9$-$2927 & \nodata & 0.075 \\
F110W\_09 & 193.22538 & $ -29.45500 $ & s, r, c & 1.24 & RDCS J1252.9$-$2927 & \nodata & 0.075 \\
F110W\_10 & 193.22505 & $ -29.45262 $ & s, r, c & 1.24 & RDCS J1252.9$-$2927 & \nodata & 0.075 \\
F110W\_11 & 193.22749 & $ -29.45127 $ & s, r, c & 1.24 & RDCS J1252.9$-$2927 & \nodata & 0.075 \\
F110W\_12 & 210.27313 & $ 2.87484 $ & p, m & 0.25 & Abell 1835 & \nodata & 0.029 \\
F110W\_13 & 210.27098 & $ 2.86989 $ & p, m & 0.25 & Abell 1835 & \nodata & 0.029 \\
F110W\_14 & 187.35722 & $ 1.84900 $ & s, j & 1.09 & \citet{santos09, dawson09} & \nodata & 0.022 \\
F110W\_15 & 218.62519 & $ 34.44785 $ & s, j & 1.24 & ISCS J1434.4+3426 & \nodata & 0.018 \\
F110W\_16 & 218.62611 & $ 34.44568 $ & s, j & 1.24 & ISCS J1434.4+3426 & \nodata & 0.018 \\
F110W\_17 & 218.62549 & $ 34.44914 $ & s, j & 1.23 & \citet{dawson09} & \nodata & 0.018 \\
F110W\_18 & 338.83679 & $ -25.96012 $ & s, r, j & 1.39 & XMMU J2235.3$-$2557 & \nodata & 0.021 \\
F110W\_19 & 338.83591 & $ -25.96062 $ & s, r, j & 1.39 & XMMU J2235.3$-$2557 & \nodata & 0.021 \\
F110W\_20 & 338.84198 & $ -25.95182 $ & s, r, j & 1.39 & XMMU J2235.3$-$2557 & \nodata & 0.021 \\
F110W\_21 & 338.83613 & $ -25.96230 $ & s, r, j & 1.39 & XMMU J2235.3$-$2557 & \nodata & 0.021 \\
F110W\_22 & 338.83593 & $ -25.96250 $ & s, r, j & 1.39 & XMMU J2235.3$-$2557 & \nodata & 0.021 \\
\cutinhead{F110W, Pre-NCS}

F110W\_61K\_01 & 209.95569 & $ 62.51310 $ & p, b, f & 0.32 & \citet{fisher98} & \nodata & 0.019 \\
F110W\_61K\_02 & 209.95828 & $ 62.51344 $ & p, b, f & 0.32 & \citet{fisher98} & \nodata & 0.019 \\
F110W\_61K\_03 & 209.95741 & $ 62.51513 $ & p, b, f & 0.33 & \citet{fisher98} & \nodata & 0.019 \\

\cutinhead{F160W, Post-NCS}

F160W\_01 & 53.07643 & $ -27.84864 $ & i, t & 1.54 & v, \citet{szokoly04} & Y & 0.007 \\
F160W\_02 & 53.06273 & $ -27.72659 $ & i, o & 1.87 & \citet{balestra10} & Y & 0.009 \\
F160W\_03 & 53.06110 & $ -27.72709 $ & i, o & 0.98 & v, \citet{lefevre04} & Y & 0.009 \\
F160W\_04 & 189.23715 & $ 62.21721 $ & y, h, w & 1.24 & q, \citet{barger08} & Y & 0.013 \\
F160W\_05 & 189.23575 & $ 62.21603 $ & y, h, w & 1.225 & q & Y & 0.013 \\
F160W\_06 & 189.22982 & $ 62.21776 $ & y, h, w & 0.95 & q, \citet{barger08} & N & 0.013 \\
F160W\_07 & 189.25714 & $ 62.20662 $ & y, h, w & 1.19 & q, \citet{barger08} & Y & 0.012 \\
F160W\_08 & 189.25511 & $ 62.20382 $ & y, h, w & 1.52 & q, \citet{cohen00} & Y & 0.012 \\
F160W\_09 & 189.03076 & $ 62.16874 $ & w, g & 0.64 & q, \citet{barger08} & N & 0.011 \\
F160W\_10 & 189.36618 & $ 62.34293 $ & x, d & 1.15 & q, \citet{wirth04} & N & 0.013 \\
F160W\_11 & 53.15855 & $ -27.69138 $ & i, o & 0.67 & v, \citet{lefevre04} & N & 0.009 \\
F160W\_12 & 53.17661 & $ -27.69827 $ & i, o & 0.68 & \citet{lefevre04} & \nodata & 0.009 \\
F160W\_13 & 53.16681 & $ -27.73859 $ & i, u & 0.52 & v, \citet{lefevre04} & N & 0.008 \\
F160W\_14 & 53.19196 & $ -27.91250 $ & d, t & 0.73 & v, \citet{vanzella08} & N & 0.008 \\
F160W\_15 & 53.18202 & $ -27.92357 $ & l, t & 0.46 & \citet{lefevre04} & \nodata & 0.007 \\
F160W\_16 & 53.13239 & $ -27.81427 $ & h, u, t & 0.77 & v, \citet{vanzella08} & N & 0.008 \\
F160W\_17 & 7.28240 & $ -0.93077 $ & k, n & 0.23 & \citet{abazajian09} & N$^2$ & 0.021 \\
F160W\_18 & 137.86492 & $ 5.85092 $ & z, e & 0.78 & \citet{kneib00} & \nodata & 0.045 \\
F160W\_19 & 137.86643 & $ 5.84706 $ & z, e & 0.76 & \citet{kneib00} & \nodata & 0.045 \\
F160W\_20 & 137.86593 & $ 5.84595 $ & z, e & 0.76 & \citet{kneib00} & \nodata & 0.045 \\
F160W\_21 & 137.86604 & $ 5.84474 $ & z, e & 0.78 & \citet{kneib00} & \nodata & 0.045 \\

\cutinhead{F160W, Pre-NCS}

F160W\_61K\_01 & 137.86492 & $ 5.85092 $ & a, z & 0.78 & \citet{kneib00} & \nodata & 0.045 \\
F160W\_61K\_02 & 137.86643 & $ 5.84706 $ & a, z & 0.76 & \citet{kneib00} & \nodata & 0.045 \\
F160W\_61K\_03 & 137.86593 & $ 5.84595 $ & a, z & 0.76 & \citet{kneib00} & \nodata & 0.045 \\
F160W\_61K\_04 & 137.86604 & $ 5.84474 $ & a, z & 0.78 & \citet{kneib00} & \nodata & 0.045 \\
F160W\_61K\_05 & 209.95617 & $ 62.51328 $ & p, b & 0.32 & \citet{fisher98} & \nodata & 0.019 \\
F160W\_61K\_06 & 209.95872 & $ 62.51361 $ & p, b & 0.32 & \citet{fisher98} & \nodata & 0.019 \\
F160W\_61K\_07 & 209.95785 & $ 62.51530 $ & p, b & 0.33 & \citet{fisher98} & \nodata & 0.019 \\

\enddata

 \tablecomments{Galaxies labeled ``61K'' are pre-NCS. The HST Program IDs are as follows: a = GO-7887, 
b = GO/DD-7941, 
c = GTO/ACS-9290, 
d = GO-9352, 
e = GO-9375, 
f = GTO/ACS-9717, 
g = GO-9856, 
h = GO-10189, 
i = GO-10258, 
j = GO-10496, 
k = GO-10886, 
l = GO-11135, 
m = GO-11143, 
n = GO-11202, 
o = GO/DD-11359, 
p = GO-11591, 
q = GO-11600, 
r = GO/DD-11799, 
s = GO/DD-12051, 
t = GO-12061, 
u = GO-12062, 
v = GO-12177, 
w = GO-12443, 
x = GO-12444, 
y = GO-12445, 
z = GO-12874.
\tablenotetext{1}{This galaxy displays IR emission lines.}
\tablenotetext{2}{This galaxy displays no optical emission lines, which, at this redshift, likely implies no IR emission lines.}   }
\end{deluxetable*}

\section{Cross-Calibration Procedure}

The ideal cross-calibration procedure would be to constrain the relative amplitude of the galaxy in each filter by directly modeling each pixel in each image, marginalizing out nuisance parameters for the underlying distribution of galaxy light on the sky, the exact alignment of the images, and the relative background levels. However, we instead selected cross-convolution for our analysis, as this approach limits the impact of systematics involved in understanding the PSF. The resulting increase in statistical uncertainty due to the convolution is limited by the convenient fact that the galaxies are significantly broader than the PSFs. (PSF systematics are suppressed to some extent when doing supernova photometry, as these systematics also affect measurements of standard stars, and only the differential measurement is important.)

We first resample the data onto the same pixel scale and orientation using \texttt{astrodrizzle} \citep{fruchter10}. In short, this package resamples individual exposures into the same (distortion-free) frame, performs an initial robust image combination, rejects discrepant pixels (in the frames of the individual exposures), and then resamples the good pixels from each individual exposure to one final combined image. The name comes from the process of resampling, in which flux in the individual image is convolved with a kernel and then ``drizzled'' into a common undistorted frame.

Using PSFs derived from bright stars, we cross-convolve the images for each filter/instrument pair to be compared, giving the same PSF for both images (technical details in Appendix~\ref{sec:dataprocessing}). Once each pair of images has the same PSF, we centroid each galaxy, then compute fluxes in annuli around that centroid (\ref{sec:instrumentalcolor}). We simultaneously fit for the true radial flux of the object (in the cross-convolved images), the relative sky level, and a scaling parameter. This scaling parameter (in units of magnitudes) represents the instrumental color of the galaxy in the pair of filters considered (instrumental color in that the zeropoints have not been taken into account). We then convert these instrumental colors to ST magnitude differences (\ref{sec:zeropointoffset}) using the \citet{brown14} galaxy templates \citep[as a cross-check, we use templates from ][]{bruzual03}. Finally, we compute the average zeropoint offsets in \ref{sec:globalzeropoint}, including remaining uncertainty in the NIC2 bandpasses (see~\ref{sec:nicmoseffectivebandpass}) and uncertainty in the NIC2 CRNL.

To prevent inadvertent bias towards the expected results, our analysis was ``blinded'' (the zeropoints were kept hidden) until the analysis was complete. The order of the unblinding was as follows. First, we checked the code on bright standard stars\footnote{HST program IDs SM2/NIC-7049, SM2/NIC-7152, CAL/NIC-7607, CAL/NIC-7691, CAL/NIC-7693, CAL/NIC-7902, CAL/NIC-7904, SM3/NIC-8983, SM3/NIC-8986, ENG/NIC-9324, CAL/NIC-9325, SNAP-9485, CAL/NIC-9639, GO-9834, CAL/NIC-9995, CAL/NIC-9997, CAL/NIC-10381, CAL/NIC-10454, GO-10496, CAL/NIC-10725, CAL/NIC-10726, CAL/NIC-11060, CAL/NIC-11061, CAL/NIC-11319, SM4/WFC3-11439, SM4/WFC3-11451, GO-11557, GO/DD-11799, CAL/WFC3-11921, CAL/WFC3-11926, GO/DD-12051, CAL/WFC3-12333, CAL/WFC3-12334, CAL/WFC3-12341, CAL/WFC3-12698, CAL/WFC3-12699, CAL/WFC3-13088, and CAL/WFC3-13089.}, ensuring that the cross-convolution code matched the results of aperture photometry on the input images (\texttt{cal}/\texttt{flt}/\texttt{flc}, see below). This is a powerful test of the PSF models, as stars are much sharper than the calibration galaxies. Then, we unblinded the F160W results, as that band is less important for the cosmological results, and could have revealed gross problems with the analysis. (We made no analysis changes after unblinding the F160W.) Finally, we unblinded the F110W. We note that only the zeropoint offsets were kept hidden; the dispersions were never hidden, and provided one avenue of feedback for the proper drizzle settings (described in \ref{sec:dataprocessing}) and the annuli-annuli correlations (described in Section~\ref{sec:anncorrelations}). The dispersion of the bright-star observations was a particularly useful diagnostic.

We evaluate the systematic uncertainties by changing assumptions one at a time (e.g., changing the minimum inner annuli radius of the photometry) and rerunning the analysis. To be conservative, the entire range (i.e., maximum $-$ minimum) is taken to be the $1\sigma$-size of that systematic. We sum these differences in quadrature.\footnote{This procedure is not optimal in the presence of heterogeneous statistical and systematic uncertainties. We test our results by computing the RMS scatter over all analyses for each galaxy, adding it in quadrature to the uncertainties for each galaxy, and refitting the mean offset. The shifts in mean offset are only 1~mmag, so the heterogeneous effects are small.} The full details of the uncertainty analysis are in Appendix~\ref{sec:uncertaintyanalysis}) while the contribution from each uncertainty is presented in Table~\ref{tab:uncertainties}. The composition of the total uncertainty depends on the calibration, but statistical uncertainty, PSF systematics, and the calibration of the galaxy templates share the bulk of it. WFC3 has its own calibration uncertainties, which we evaluate in~\ref{sec:WFCuncertainties}. These are currently comparable to the cross-calibration uncertainties, but may be reduced with future calibration programs.

\begin{deluxetable*}{lrrrr}
\tablecolumns{5}
\tabletypesize{\scriptsize}
\tablecaption{Uncertainties present in the cross-calibrations.}
\tablehead{
 \colhead{Uncertainty} &
 \colhead{F110W} &
 \colhead{F160W} &
 \colhead{pre-NCS F110W} &
 \colhead{pre-NCS F160W}
}
\startdata
Statistical & \ErrorJStat & \ErrorHStat & \ErrorJpreStat & \ErrorHpreStat \\
Calibration of Color-Color & \ErrorJColCol & \ErrorHColCol  & \ErrorJpreColCol & \ErrorHpreColCol\\
Encircled Energy Correction & \ErrorJEE & \ErrorHEE & \ErrorJpreEE & \ErrorHpreEE \\
PSF Shape & \ErrorJPSF & \ErrorHPSF & \ErrorJprePSF & \ErrorHprePSF \\
NICMOS Effective Bandpass & \ErrorJWave & \ErrorHWave & \ErrorJWave & \ErrorHWave \\
Annuli Correlations & \ErrorJAnn & \ErrorHAnn & \ErrorJpreAnn & \ErrorHpreAnn \\
Templates and Extinction & \ErrorJTemTotal & \ErrorHTemTotal & \ErrorJpreTemTotal &  \ErrorHpreTemTotal\\
\hline
Total & \ErrorJTotal &  \ErrorHTotal & \ErrorJpreTotal &  \ErrorHpreTotal 
\enddata
\label{tab:uncertainties}
\end{deluxetable*}

\section{Results}

\begin{deluxetable*}{lccc}
\tablecolumns{4}
\tabletypesize{\scriptsize}
\setlength{\tabcolsep}{0.02in}
\tablecaption{The results of our measurements.}

\tablehead{
 \colhead{Fit} &
 \colhead{NIC2 ST Zeropoint} &
 \colhead{NIC2 Low-Count-} &
 \colhead{STScI ST Zeropoint}\\
 & \colhead{$-$ WFC3 ST Zeropoint} & \colhead{Rate ST Zeropoint} & \\
}
\startdata
\cutinhead{F110W}
{\bf WFC3 \Suggested Bandpass} & \JZPOffsetrev & \NICJSTZPrev & \stsciJSTZPpostNCS \\
\Given Bandpass &                           \JZPOffsetsyn & \NICJSTZPsyn & \stsciJSTZPpostNCS\\
{\bf Pre-NCS, WFC3 \Suggested Bandpass} & \JpreZPOffsetrev & \NICJpreSTZPrev & \stsciJSTZPpreNCS \\
Pre-NCS, \Given Bandpass &           \JpreZPOffsetsyn & \NICJpreSTZPsyn & \stsciJSTZPpreNCS \\
\cutinhead{F160W}
{\bf WFC3 \Suggested Bandpass} & \HZPOffsetrev & \NICHSTZPrev & \stsciHSTZPpostNCS \\
\Given Bandpass &                           \HZPOffsetsyn & \NICHSTZPsyn & \stsciHSTZPpostNCS \\
{\bf Pre-NCS, WFC3 \Suggested Bandpass} & \HpreZPOffsetrev & \NICHpreSTZPrev & \stsciHSTZPpreNCS \\
Pre-NCS, \Given Bandpass &           \HpreZPOffsetsyn & \NICHpreSTZPsyn & \stsciHSTZPpreNCS \\
\enddata
\label{tab:results}
 \tablecomments{The bolded items are the recommended results. The difference in zeropoints is the quantity $k_0^{\mathrm{ST}}$, described in Section~\ref{sec:globalzeropoint}.}
\end{deluxetable*}

\begin{figure}[htp]
\tikzstyle{startstop} = [rectangle, rounded corners, minimum width=5cm, minimum height=1cm,text centered, draw=black, text width=5cm]
\tikzstyle{arrow} = [thick,-]
\tikzstyle{between} = [rectangle, minimum width=5cm, minimum height=0.5cm,text centered, text width=5cm]

\begin{center}
\begin{tikzpicture}[node distance=1.0 cm]

\node (wfc3bluehigh) [startstop] {WFC3 Blue/CALSPEC Zeropoints, High-Count-Rate Calibration};
\node (wfc3bandpass) [between, below of=wfc3bluehigh] {WFC3 Bandpass, Section~\ref{sec:wfceffectivebandpass}};

\node (wfc3sthigh) [startstop, below of=wfc3bandpass] {WFC3 ST Zeropoint, High-Count-Rate Calibration};
\node (wfc3crnl) [between, below of=wfc3sthigh] {WFC3 CRNL Measurements};
\node (wfc3stlow) [startstop, below of=wfc3crnl] {WFC3 ST Zeropoint, Low-Count-Rate Calibration};
\node (galaxy) [between, below of=wfc3stlow] {Galaxy Observations, described in this work};
\node (nic2stlow) [startstop, below of=galaxy] {NIC2 ST Zeropoint, Low-Count-Rate Calibration};
\node (nic2bandpass) [between, below of=nic2stlow] {NIC2 Bandpass Model, Section~\ref{sec:nicmoseffectivebandpass}};
\node (nic2bluelow) [startstop, below of=nic2bandpass] {NIC2 Blue/CALSPEC Zeropoints, Low-Count-Rate Calibration};

\draw [arrow] (wfc3bluehigh) -- node { } (wfc3bandpass);
\draw [arrow] (wfc3bandpass) -- node { } (wfc3sthigh);

\draw [arrow] (wfc3sthigh) -- node { } (wfc3crnl);
\draw [arrow] (wfc3crnl) -- node { } (wfc3stlow);

\draw [arrow] (wfc3stlow) -- node { } (galaxy);
\draw [arrow] (galaxy) -- node { } (nic2stlow);

\draw [arrow] (nic2stlow) -- node { } (nic2bandpass);
\draw [arrow] (nic2bandpass) -- node { } (nic2bluelow);

\end{tikzpicture}
\end{center}

\caption{Diagram for the paths between the zeropoints discussed in this work. The ``Blue/CALSPEC'' zeropoints are referenced to CALSPEC calibration stars like solar analogs and Vega, which are bluer than the ST magnitude reference (flat in $f_{\lambda}$). The WFC3 CRNL measurements come from \citet{riess10a, riess10b, riess11}.}\label{fig:calsummary}

\end{figure}
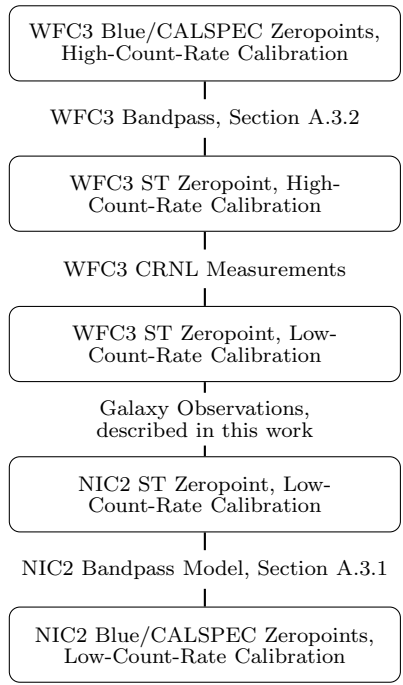

Table~\ref{tab:results} presents our ST magnitude zeropoint differences for both the \given and \suggested WFC3 bandpasses (WFC3 bandpasses are discussed in \ref{sec:wfceffectivebandpass}). We recommend the analyses highlighted in boldface type; any revisions to the WFC3 bandpasses can be interpolated from the pair of numbers from each result. Likewise, any updates to the understanding of the WFC3 zeropoints at low count-rates can be propagated through the results presented here into the NIC2 zeropoints. (The correlations between NICMOS and WFC3 zeropoints specified here should be taken into account in cosmological fits using the SN Ia Hubble diagram with both NIC2 and WFC3-observed SNe.)

Here, we illustrate the application of these results to the NIC2 zeropoints at low count-rates, applicable to any NIC2 \texttt{cal} files processed with the steps in \ref{sec:dataprocessing}. We note that our zeropoints assume 1\arcsec-radius encircled energy corrections for F110W of \NICJEE and \NICHEE for F160W (we use PSF photometry for the supernova data, but the PSFs are normalized to these values), discussed further in \ref{sec:dataprocessing}.

Figure~\ref{fig:calsummary} summarizes the paths for moving between the zeropoints we reference. Our calibration cross-calibrates WFC3 and NIC2, so we start with the WFC3 zeropoints. The observed Vega WFC3 F110W high-count-rate zeropoint (with our suggested bandpass revision) is 
26.072. Accounting for the WFC3 CRNL, the zeropoint at low count-rates is 26.032. Converting to ST magnitude (using our bandpass revision) gives an ST zeropoint of 28.434. Applying our cross-calibration gives 25.296 for NIC2 ST. This same sequence was applied to both F110W and F160W; the resulting zeropoints are shown in second column of column of Table~\ref{tab:results}. For comparison, we take the NIC2 Vega STScI zeropoints, and convert to ST zeropoints using the low-count-rate conversions in Table~\ref{tab:synthzeropointdiff}. We follow this process, rather than using the STScI ST zeropoints, as the NIC2 Vega-to-ST conversion will depend on count-rate. These results are in the final column of Table~\ref{tab:results}. Our zeropoints range between 0.004 fainter (higher) for post-NCS F160W to 0.034 magnitudes fainter (higher) for post-NCS F110W, but show reasonable consistency. Other low-count-rate zeropoints (Vega or AB) can be computed using the low-count-rate offsets given in Table~\ref{tab:synthzeropointdiff}. We remind the reader that interpreting the photometric measurements should be done using a modified bandpass, as the CRNL preferentially affects blue wavelengths (discussed in \ref{sec:nicmoseffectivebandpass}).

As a modest related result, we also note that the galaxy-galaxy scatter in the zeropoint estimates is a few percent. This limits spatial variation in the NIC2 CRNL to \spatialCRL, at least on $\sim 1\arcsec$ scales.

\section{Conclusions}

This work presents a cross-calibration of the NIC2/WFC3 F110W and F160W zeropoints at the the low count-rates applicable to high-redshift SNe~Ia observations. These measurements are in tension with both the \citet{mobasher0503} results (at least 0.1 mag tension), and some earlier unpublished SCP work (0.03 magnitudes tension). We note that this tension is not due to the version of \texttt{calnica}; we get essentially the same NICMOS magnitudes with the improved version 4.4.1 as with the older 4.1.1 that the pre-2008 results were run with \citep[see the discussion of the improvements in][]{dahlen0802}. Our results show no tension with the higher-count-rate zeropoint and CRNL measurements, with our results having smaller uncertainties at low count rates. A new ``Union'' compilation of SNe using this calibration will be presented in a future paper.

\acknowledgements{
Financial support for this work was provided by NASA through programs GO/DD-11799 and GO/DD-12051 from the Space Telescope Science Institute, which is operated by AURA, Inc., under NASA contract NAS 5-26555. This work was also partially supported by the Director, Office of Science, Department of Energy, under grant DE-AC02-05CH11231. This research has made use of the NASA/IPAC Extragalactic Database (NED) which is operated by the Jet Propulsion Laboratory, California Institute of Technology, under contract with the National Aeronautics and Space Administration. We thank the anonymous referee for their feedback in improving this work.

{\it Facilities:} \facility{Hubble Space Telescope}.

\appendix
\section{Details of the Cross-Calibration}

\subsection{Data Processing} \label{sec:dataprocessing}

We started with the NICMOS \texttt{cal} files, flat-fielded files that have had cosmic rays rejected using the ``up-the-ramp'' multiple readouts. We processed each \texttt{cal} file first with the STSDAS task \texttt{pedsky} \citep{bushouse00}, to remove the variable quadrants seen in NICMOS data. We then ran \texttt{rnlincor}, to correct the images for the CRNL as measured at high count rates. After this processing, amplifier glow and other forms of spatially-variable background remained, so we ran the subtraction detailed in \citet{hsiao10}. We used either the ``low'' or ``high'' background models, selecting the one that minimized the median absolute deviation of the image.\footnote{This order, \texttt{rnlincor} then sky-subtraction, was the opposite order of what \citet{suzuki12, rubin13} did. The resulting difference in the supernova fluxes is only about 1\%, and we will publish an update in a forthcoming paper. Our order here seems to improve the agreement between NIC2 and WFC3 at the lowest count-rates.} We masked the erratic middle column, rather than attempting to recover the flux, as this data is far less important for our extended objects than for the SN data with which that work was concerned. Even after \texttt{pedsky} and our sky subtraction, the sky level in each image varies spatially. We thus fit for the residual sky under each galaxy, as shown in Equation~\ref{eq:residual}.

We did no pre-drizzle processing of the WFC3 (\texttt{flt}, calibrated flat-fielded exposures) data or the ACS (\texttt{flc}, calibrated, flat-fielded, charge-transfer inefficiency-corrected exposures) data except using \texttt{tweakreg} to align the images. To prevent cosmic ray hits in the ACS images from being considered objects, we aligned L.A.Cosmic \citep{vandokkum01} cleaned images. (Many of the ACS F775W visits had only one exposure per set of guide stars, so we chose to always align the input images (\texttt{flt}/\texttt{flc}), rather than stacking all of the data with a given set of guide stars and aligning the stacks.) To assist with the WFC3 alignment, we replaced each bad pixel with the median of the surrounding values. We then transferred the alignment to the original \texttt{flt} or \texttt{flc} images. For each instrument, we selected an optimal reference image based on depth and overlap with other images.

We used \texttt{astrodrizzle} to resample all data to a common pixel scale (0\farcs05, the native scale of ACS) and orientation (arbitrarily chosen to be North-up East-left) for cross-convolution. We selected a Gaussian kernel, with \texttt{pixfrac}=1 (the FWHM of the kernel in the input pixel scale). In testing, the kernel settings only had a mild impact on the dispersion of measured magnitudes. To prevent the loss of flux in the cores of bright stars, we weight each pixel in the drizzling by the exposure time of the image (this keeps the Poisson-dominated pixels from being deweighted).\footnote{For similar reasons, we also increase the minimum cosmic-ray-rejection threshold to 3.0/2.0 times the derivative (instead of the default 1.5/0.7), used with bad-pixel rejection algorithm, \texttt{minmed}. Before drizzling, we also scale the NICMOS image uncertainties by a constant factor for each image to achieve accurate uncertainties; see \citet{suzuki12} for details.} In our processing, we included some data quality (DQ) values that are non-zero, but still indicate a reliable flux measurement.\footnote{For NICMOS, we allowed pixels containing flags 512 (cosmic ray in up-the-ramp sampling), 1024 (pixel contains source), and 2048 (signal in zeroth read). For WFC3 IR, we allowed flags 2048 (signal in zeroth read) and 8192 (cosmic ray detected in up-the-ramp sampling).} Oddly, the post-NCS NICMOS data showed a difference in fluxes before and after \texttt{astrodrizzle} of 0.7\%.\footnote{This scale is such that the post-NCS \texttt{drz} images had to be scaled by 1.007 to match the \texttt{cal} images. None of the other data showed the same effect after accounting for pixel-area variation. We verified using \texttt{drizzlepac pixtosky.xy2rd} that the difference was not due to an assumed plate scale change. As we perform the supernova photometry on the \texttt{cal} images, rescaling the \texttt{drz} images is the correct procedure.}

In order to cross-convolve the images, we must have an accurate PSF for each filter. Even if we had perfect model PSFs for the observed pixels, drizzling the data onto a new set of pixels will broaden the PSFs, making empirical PSFs a necessity. To derive NICMOS and WFC3 PSFs, we downloaded P330E data (a solar-analog calibration star with many observations), and derived a convolution kernel that matches Tiny Tim \citep{krist93, krist11} PSFs to the drizzled P330E data. Although P330E is not as red as the calibration galaxies (and will thus have a slightly different PSF), it is the reddest standard for which a large amount of IR data exists. We discard any images that have non-zero DQ flags near the PSF core except the flags in the footnote. To derive ACS PSFs, we used bright, isolated stars selected from the fields, as there were not enough P330E images to derive PSFs.

These PSFs must be normalized. For this purpose, we normalize with a circular aperture of 1\arcsec-radius, which is large enough that the variation in encircled energy (EE) with object SED is a few mmag for all filters. It is also large enough to ensure that resampling the image does not affect the EE values. The normalization values for ACS F775W/F814W, taken from \citet{sirianni05}, are 0.955. For NICMOS, we compute the EE values using Tiny Tim (version 7.5) with a range of SN, galaxy, and standard-star SEDs. We use 7x oversampling ($\sim$~0\farcs01 per pixel), which matches the PSFs we use for SN photometry. (The EE values change coherently by $\sim$~0.2\% if we use 10x oversampling instead.) The average normalization values are \NICJEE for F110W and \NICHEE for F160W. For WFC3, we use the values from \citet{hartig09} (to best match the STScI WFC3 calibration): F110W: 0.932, F125W: 0.927, and F160W: 0.915.

\subsection{Fitting the Instrumental Colors, $k$} \label{sec:instrumentalcolor}
\newcommand{\multiplysymbol}{\;}

We centroid each galaxy in each drizzled stack by maximizing the flux inside a 0\farcs15 radius aperture (it makes virtually no difference if 0\farcs1 is used instead). (The signal-to-noises of these galaxies are high enough that this procedure is not significantly biased.) We then extract annular fluxes, $\mathbf{f}$, in 1-pixel-radius steps from 1 (or 3) to 10 (or 15) pixels, weighting each pixel by the fraction covered by the annulus. To obtain each color, we minimize the following expression: 
\begin{equation} \label{eq:negtwoLL}
\mathbf{r^T} \cdot C^{-1} \cdot \mathbf{r} + \log{|C|}\;.
\end{equation}
$\mathbf{r}$ is the residual from the model:
\begin{equation} \label{eq:residual}
\mathbf{r} = \mathbf{f} - [10^{-0.4 \multiplysymbol k} \multiplysymbol \mathbf{F} + s \multiplysymbol \mathbf{a}]\;,
\end{equation}
where $\mathbf{F}$ is the modeled flux of the galaxy in each annulus, $s$ is the modeled sky value, $\mathbf{a}$ is the area of each annulus, and $k$ is the modeled ratio of instrumental count-rates (measured in magnitudes). This is the count-rate ratio (as observed) between two filters and/or instruments, without correcting for the object SED or the zeropoints. There are arbitrary scaling and offset factors, which we handle by fixing $s$ and $k$ to zero for one filter. Although there is only one sky parameter, the symmetry of the annuli implies that the fit is insensitive to linear spatial variation of the sky (as well as a constant offset).

$C$ is the covariance matrix of the $\mathbf{f}$ values. The diagonal terms of $C$ are:
\begin{equation} \label{eq:Cii}
C_{ii} = \frac{10^{-0.4 \multiplysymbol k} \multiplysymbol \mathbf{F}_i + s \multiplysymbol \mathbf{a}_i}{g \multiplysymbol t} + \mathbf{v}_{\mathrm{sky}_i} \;,
\end{equation}
where $g$ is the gain of the image (ADU/electron), $t$ is the exposure time, and $\mathbf{v}_{\mathrm{sky}}$ is the sky variance as determined empirically from object-free regions of the image. The first term is the Poisson uncertainty on the count-rates of the galaxy, while the second represents sky noise. As every image gets resampled by \texttt{astrodrizzle}, then convolved with another PSF, and then integrated in annuli (which share fractional pixels between neighboring annuli), there are large off-diagonal correlations. These correlations ($\rho_{ij}$) are also found empirically from object-free regions; we then set $C_{ij} = \rho_{ij} \sqrt{C_{ii} C_{jj}}$. 

\subsection{Fitting Zeropoint Offsets, $k^{\mathrm{ST}}$, for Each Galaxy} \label{sec:zeropointoffset}

After obtaining the $k$ values (fitting out the $\mathbf{F}$ values and the $s$ values), we can fit the inter-calibrations. For the abscissa $k$ values, we scale out the following zeropoints: ACS F775W: 26.41699, ACS F814W: 26.79887, WFC3 F110W: 28.40001, WFC3 F125W: 27.9803, and WFC3 F160W: 28.1475 (these are the STScI zeropoints, with the WFC3 zeropoints shifted by 0.04 magnitudes for the WFC3 CRNL, see Section~\ref{sec:WFCuncertainties}). (The impact of the uncertainties on colors is discussed in~\ref{sec:systOtherCal}. Note that only differences in these zeropoints are meaningful, as they are used only to measure the color of each galaxy.) This gives us the abscissa ST magnitude color 
for each galaxy in the analysis. We fit linear relations to the color-color relations (see Figure~\ref{fig:colorcolor} for typical relations), using templates with abscissa ST magnitudes within 0.25 of each observed galaxy. (This $\pm 0.25$ magnitude cut ensures that our results are not affected by the fact that the relations are not quite linear. This cut is large enough such that we always have several templates available to derive a local relation.) We subtract these relations from the ordinal $k$ values, producing estimates of the ST magnitude difference between NICMOS and WFC3, $k^{\mathrm{ST}}$.

We use the \citet{brown14} galaxy templates for our primary analysis. These templates are constructed using spectra and photometry of 129 nearby galaxies, with some interpolation using (mostly) stellar population synthesis models from \citet{bruzual03} (plus dust and PAH components). Although constructed from nearby galaxies, they match observed color-color relations at $z\sim0.4$ \citep[for details, see][]{brown14}, lending support to their use at higher redshift. As a cross-check, we use \citet{bruzual03} models, but do not use this as our primary analysis.

Synthesizing the color-color relations requires knowledge of the bandpasses, especially of NICMOS and WFC3 F110W and F160W (because of the shallow slopes of the color-color relations, the other bands are less important). Uncertainties in these bandpasses are described below.

\subsubsection{NICMOS Effective Bandpass} \label{sec:nicmoseffectivebandpass}

The NICMOS CRNL depends strongly on wavelength, so the effective bandpasses of NICMOS will depend on count-rate, as illustrated in Figure \ref{fig:effectivebandpassplot}. We measure excellent agreement between synthesized (using the 2014 March CALSPEC\footnote{\url{http://www.stsci.edu/hst/observatory/crds/calspec.html}}) and measured magnitudes among G191-B2B, GD153, GD71, GRW+70 5824, WD1657+343, P041C, P177D, P330E, SNAP-2, VB8 (the data here are saturated in F160W), 2M0036+18, and 2M0559-14 (F110W data only) using the \texttt{synphot} NIC2 bandpasses at high count-rates. This check limits any significant deviation from the \given bandpass to only the effects of the CRNL. There are no blue NIC2 medium or narrow-band filters, and no NIC2 grism, so we we cannot measure the change in NIC2 CRNL with wavelength. However, the CRNL (in mag/dex) is roughly linear with wavelength for NIC3, where it was measured in small wavelength bins using the grisms. We thus parameterize the effect on the NIC2 bandpasses using a function that is linear (in magnitudes) with respect to wavelength (i.e., an $\exp{(\lambda)}$ bandpass warping function). This function is constrained to be 0.063 magnitudes/dex at 11,000\ang, and 0.029 magnitudes/dex at 16,000\ang, matching the high-count-rate \citet{dejong0601} measurements of the NIC2 CRNL in the F110W and F160W data, respectively. As we do not know the effective wavelength of amplifier glow (or how to treat dark current), we do not assume that this function should be evaluated with four dex (for the four dex separating the supernovae and standards). We instead parameterize the deviation from the high-count-rate bandpass in terms of the nuisance parameter $\beta$ (see Section~\ref{sec:globalzeropoint}), which warps the bandpasses by $10^{-\beta \frac{2}{5} 0.063}$ at 11,000\ang and $10^{-\beta \frac{2}{5} 0.029}$ at 16,000\ang. Our standard analysis conservatively assumes a Gaussian prior of $2 \pm 2$ on $\beta$, so that both four and zero are easily accommodated.

The sensitivity of NICMOS improved preferentially in the blue with the installation of the NCS. For the pre-NCS data, the bandpass must therefore be adjusted. Turning again to the NIC3 grism data (in G096L and G141L), we see that the pre/post-NCS sensitivity change is roughly linear with wavelength. As with the wavelength dependence of the CRNL, we fix the NIC2 slope with wavelength using the pre/post-NCS zeropoint change in the F110W and the F160W (0.45 and 0.33 magnitudes\footnote{
{\footnotesize \url{http://www.stsci.edu/hst/nicmos/performance/photometry/postncs_keywords.html}} and {
\footnotesize \url{http://www.stsci.edu/hst/nicmos/performance/photometry/prencs_keywords.html}}}). This lets us handle pre-NCS data with the same bandpass model, just with the above prior on $\beta$ changed to $5.6 \pm 2$.

\begin{figure}[h]
\begin{center}
\includegraphics[width=3.2in]{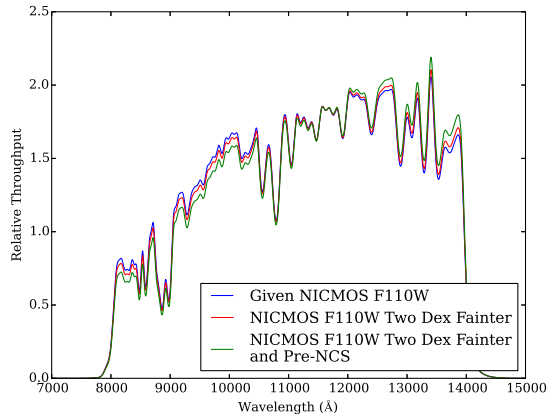}
\end{center}
\caption{Plots of our model of the NIC2 F110W bandpass under different conditions, with arbitrary normalization. The blue line shows the bandpass taken from \texttt{synphot}. Observations of bright standard stars with a range of colors show consistency with this bandpass at those high count-rates. In red, we show the assumed bandpass two dex fainter ($\beta=2$). As described in Section \ref{sec:nicmoseffectivebandpass}, the CRNL preferentially acts at blue wavelengths, shifting the bandpass effective wavelength to the red for lower count-rates. The pre-NCS NIC2 had worse sensitivity at blue wavelengths, giving a further effective-wavelength shift to the red shown in green ($\beta=5.6$).
\label{fig:effectivebandpassplot}}
\end{figure}

\subsubsection{WFC3 Effective Bandpass} \label{sec:wfceffectivebandpass}

Unlike NICMOS, the WFC3 CRNL is roughly independent of wavelength. Thus, establishing the WFC3 bandpasses at high count-rates is sufficient for all count-rates. (Although the galaxies in this analysis are close to zero ST color on average (flat in $f_{\lambda}$), knowledge of the WFC3 bandpasses is necessary to compute the ST magnitude zeropoint from the bluer standard stars.) As with NIC2, we check the observed and synthesized magnitudes of the standard stars G191-B2B, GD153, GD71, GRW+70 5824, WD1657+343, P041C, P177D, P330E, SNAP-2, and KF06T2 (for F160W, there is also data for VB8). These stars span a smaller range of colors than the stars observed with NIC2, but strongly indicate that shifts of the bandpasses to the red are necessary. Coincidentally, the effective-wavelength shifts needed for both filters are 60\ang. We implement these shifts using the same smooth warping function used in Section~\ref{sec:nicmoseffectivebandpass}. As we only need the bandpass for converting between the Calspec-derived zeropoints and the ST magnitude zeropoints, the choice of functional form for the effective-wavelength shift will only have a small effect.

\subsubsection{Synthesized High-Count-Rate Zeropoints}

In Table~\ref{tab:highcountzeropoints}, we present our Vega zeropoints derived from standard stars using 1\arcsec-radius aperture photometry. (Vega is close in color to the average standard used in this determination; these zeropoints can be transformed using Table~\ref{tab:synthzeropointdiff}.) The WFC3 bright zeropoints are fainter than the STScI zeropoints\footnote{\url{http://www.stsci.edu/hst/wfc3/phot_zp_lbn}}, as noted by \citet{nordin14} (who used PSF photometry). The WFC3 Vega zeropoints are almost independent of bandpass used, as the average color of the standard stars is not very dissimilar from Vega. Varying the photometry radius used can vary the zeropoints by several mmag, so these zeropoints are only tied to CALSPEC at the level of $\sim~0.01$ mag. The CALSPEC system itself also has uncertainty, so all of these zeropoints are most accurately defined with respect to other calibrations to that system. The NICMOS zeropoints show good agreement with their STScI counterparts, with some scatter. We note that the NICMOS bright zeropoints are only presented for comparison to the faint zeropoints, and do not enter our analysis (except to constrain the pre-NCS NIC2 bandpasses).

\begin{deluxetable}{lcc}
\tablecolumns{3}
\tabletypesize{\footnotesize}
 \tablecaption{High-Count-Rate Vega Zeropoints
 \label{tab:highcountzeropoints}}
 \tablehead{
 \colhead{Bandpass} & \colhead{Observed Zeropoint} & \colhead{STScI Zeropoint}}
 \startdata 
WFC3 F110W, Synphot & 26.074 & 26.063 \\
WFC3 F110W, Suggested Revision & 26.072 & \nodata  \\
WFC3 F160W, Synphot & 24.708  &  24.695  \\
WFC3 F160W, Suggested Revision & 24.708 &  \nodata \\
NICMOS F110W, Synphot & 22.973 &  22.964  \\
NICMOS F110W, Pre-NCS & 22.500 &  22.500  \\
NICMOS F160W, Synphot & 22.144 & 22.153  \\
NICMOS F160W, Pre-NCS & 21.816 &  21.816
 \enddata
 \tablecomments{These zeropoints are computed using 1\arcsec-radius aperture photometry of bright standards with the March 2014 CALSPEC spectra. The proposed revisions of the WFC3 bandpasses have little effect on the Vega zeropoints, as the average color of the standards is not far from Vega. We find fainter (larger) zeropoints for WFC3, in accordance with \citet{nordin14}. Our high-flux NICMOS zeropoints are presented for comparison only, and do not enter our analysis.}
\end{deluxetable}

\begin{deluxetable}{lccc}
\tabletypesize{\footnotesize}
\tablecolumns{4} 
 \tablecaption{Synthesized Zeropoint Differences
 \label{tab:synthzeropointdiff}}
 \tablehead{
 \colhead{Bandpass} & \colhead{Effective} & \colhead{ST $-$ Vega} & \colhead{AB $-$ Vega} \\
& \colhead{Wavelength} & \colhead{(Mag)} & \colhead{(Mag)}
 }
 \startdata 

WFC3 F110W, Synphot & 11797 &  2.3826 &  0.7647 \\
WFC3 F110W, Suggested Revision & 11857 &  2.4024 &  0.7728 \\
WFC3 F160W, Synphot & 15436 &  3.4978 &  1.2566 \\
WFC3 F160W, Suggested Revision & 15496 &  3.5131 &  1.2634 \\
NICMOS F110W, Synphot & 11575 &  2.2936 &  0.7328 \\
NICMOS F110W, Suggested Low-CR & 11605 &  2.3036 &  0.7366 \\
NICMOS F110W, Pre-NCS and Low-CR & 11659 &  2.3215 &  0.7434 \\
NICMOS F160W, Synphot & 16159 &  3.6474 &  1.3147 \\
NICMOS F160W, Suggested Low-CR & 16175 &  3.6515 &  1.3165 \\
NICMOS F160W, Pre-NCS and Low-CR & 16206 &  3.6588 &  1.3196 \\

 \enddata
 \tablecomments{This table is intended to aid conversions among the different magnitude systems. We present the effective wavelength of each filter, computed for a source flat in $f_{\lambda}$. We also present ST $-$ Vega and AB $-$ Vega magnitude conversions. Each WFC3 result is presented with and without our proposed bandpass shift. We also present results using the NIC2 bandpasses at high-count rates, with the effects of the CRNL taken into account for low count-rates, and pre-NCS at low-count rates.}
\end{deluxetable}

\subsection{Fitting the Global Zeropoint Differences, $k_0^{\mathrm{ST}}$} \label{sec:globalzeropoint}

Tests involving fitting a scale between images of the same galaxies in WFC3 data (with a range of spatial offsets and rotations) reveals a $\sim~\WFCWFCscatter$ magnitude scatter. We take this as due to different pixel sampling in the undersampled images. The existence of this irreducible scatter implies that the statistical uncertainty is best judged (in part) using the observed dispersion of the scale factors about the mean. We must take into account residual uncorrected CRNL for both NIC2 and WFC3, as well as the partially known effective bandpass at these low count-rates. As we have enough data points to reliably estimate both calibration parameters and uncertainties using maximum likelihood, we minimize the following expression for each calibration: 
\begin{equation}
\sum_{i} \frac{[k^{\mathrm{ST}}_i - (k_0^{\mathrm{ST}} + \alpha \multiplysymbol [M_i^{\mathrm{rnlincor}} - M_{\mathrm{mean\ SN}}^{\mathrm{rnlincor}}] + \beta \multiplysymbol C_i^{\mathrm{rnlincor}})]^2}{\sigma^2_i + \sigma_{\mathrm{int}}^2}  +  \sum_i \log(\sigma^2_i + \sigma_{\mathrm{int}}^2)\;.
\label{eq:meancalibration}
\end{equation}

\noindent $k_i^{\mathrm{ST}}$ is the ST magnitude NIC2$-$WFC3 difference measurement for each galaxy (with measurement uncertainty $\sigma^2_i$). $M_i^{\mathrm{rnlincor}}$ is the amount of non-linearity correction \texttt{rnlincor} applies to each galaxy. It is thus a surrogate count-rate measurement, with lower count-rates giving higher corrections. $M_{\mathrm{mean\ SN}}^{\mathrm{rnlincor}}$ is the mean \texttt{rnlincor} correction for the high-redshift SNe~Ia near maximum, equal to \meanJnlc in F110W and \meanHnlc for F160W. $C_i^{\mathrm{rnlincor}}$ is the change in $k_i^{\mathrm{ST}}$ with respect to a change in count-rates by one dex due to the estimated wavelength-dependence of the CRNL (with respect to ST magnitude); it is thus a measure of the color of each galaxy. (Emission lines also play a role, but most of the variation is due to color.) We present our measurements of these parameters in Table~\ref{tab:calib}.

The fit parameters are as follows. $k_0^{\mathrm{ST}}$ is the zeropoint offset defined for zero ST color and $M_{\mathrm{mean\ SN}}^{\mathrm{rnlincor}}$. $\alpha$ parameterizes any residual CRNL in either NIC2 or WFC3 (for simplicity, we assume that the WFC3 CRNL is proportional to the NIC2 CRNL). As described in Section \ref{sec:nicmoseffectivebandpass}, $\beta$ is used to measure any deviation from the \given high-count-rate bandpasses. Finally, $\sigma_{\mathrm{int}}^2$ is a fit parameter representing irreducible variance (assumed to be the same for all galaxies in one band). The sum is usually over each galaxy. As there are not enough objects in the small NICMOS FoV to align separate NICMOS datasets, the sum ranges over these datasets, if more than one is present for a galaxy. As discussed in Section~\ref{sec:nicmoseffectivebandpass}, we take a prior of $2 \pm 2$ on $\beta$ for the post-NCS NICMOS data, and $5.6 \pm 2$ for the pre-NCS data. For the post-NCS data, we do not take any prior on $\alpha$, as we are testing for deviation from the predicted low-count-rate behavior. For the pre-NCS data (which uses many fewer objects), we assume that the WFC3 CRNL is 0.01 mag/dex, and the NICMOS CRNL is adequately corrected over this narrow range of count-rates (as it seems to have been in this count-rate range for the post-NCS data). $\alpha$ is thus fixed to 0.01 mag/dex / 0.063 mag/dex = 0.1587 for the pre-NCS F110W data and 0.01/0.029 = 0.3448 for the pre-NCS F160W data \citep[recall that the 0.063 and 0.029 come from the][measurements of the NIC2 CRNL at high count-rates]{dejong0601}.

Illustrations of the fits are shown in Figure~\ref{fig:kST}. We note that for the F110W data, it appears that the NIC2$-$WFC3 zeropoint gap narrows at very low count-rates (visible as higher points towards the right in the left panels). It may be that \texttt{rnlincor} over-corrects NIC2 F110W at these count-rates. Additional systematic uncertainty is likely called for when using \texttt{rnlincor} corrections greater than 0.25 magnitudes for NIC2 F110W.

There are three faint stars in the F110W data, allowing us to use them as a cross-check. For these, we use the \citet{pickles98} stellar library for the color-color relation. As with the pre-NCS data, we fix $\alpha$, as we do not have enough objects over a large enough range of count-rates to reliably fit it. Large-aperture photometry on faint stars does not give high signal-to-noise, but we do find consistency with the galaxy results: $k_0^{\mathrm{ST}} = $ \StarsJ using the revised WFC3 bandpass.

As another cross-check, we fix $\alpha$ for the post-NCS data to investigate how fitting out uncorrected CRNL affects our results. The calibrations, using the modified WFC3 bandpasses, are only \Jfixalphadiff and \Hfixalphadiff (F110W and F160W, respectively) different. These tests indicate that our mean galaxy count-rate is close to the mean supernova count-rate. As a similar cross-check, we have objects spanning enough of a color range in F110W to unfix $\beta$ (although we now fix $\alpha$ for maximum statistical power). This results in a zeropoint difference of \Jfitbetadiff. Encouragingly, we find a $\beta$ measurement of \Jfitbetabeta, more consistent than not with the need to modify the bandpass at lower count rates.

\begin{figure*}[h]

\includegraphics[width=0.5\textwidth]{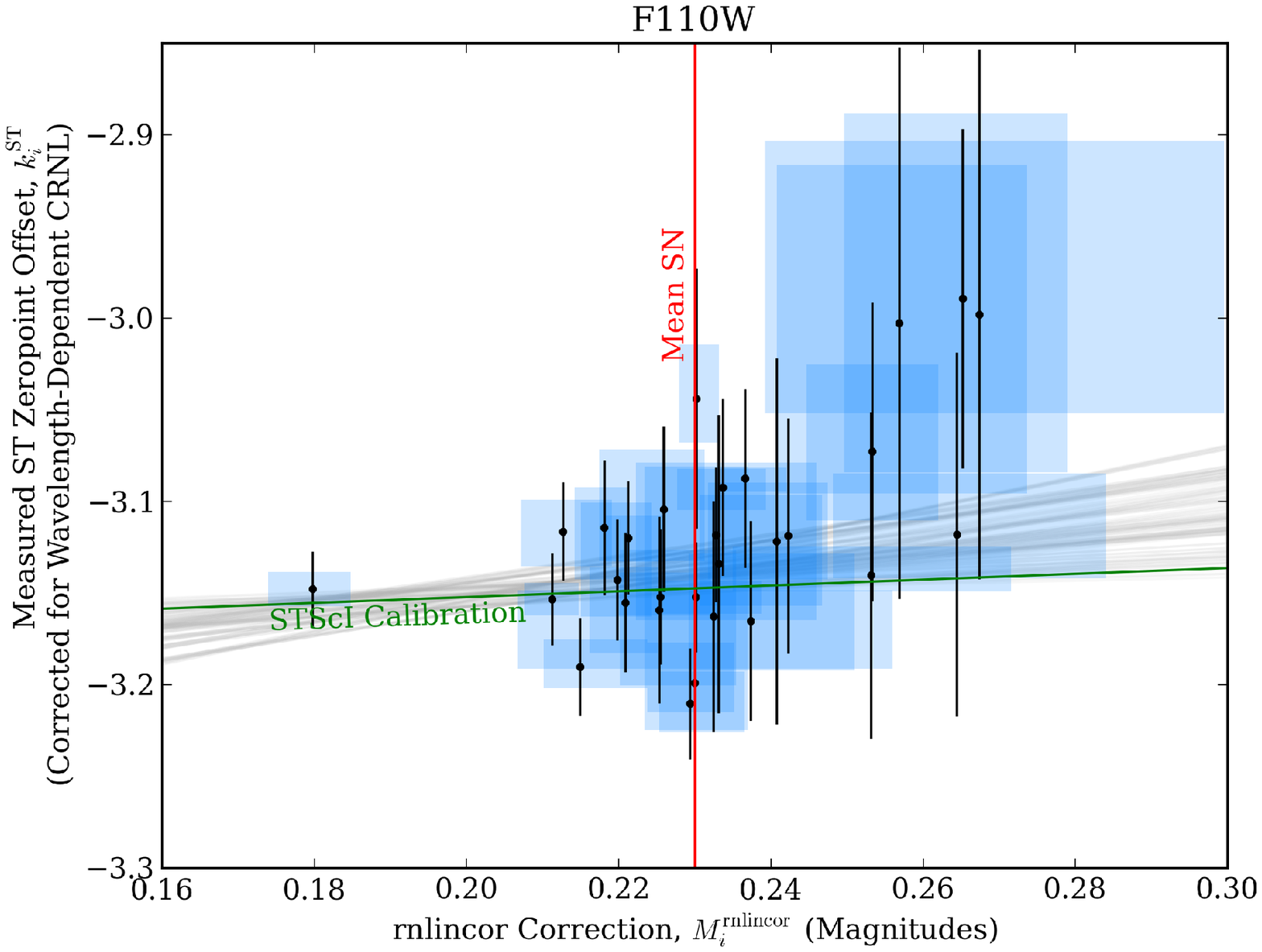}
\includegraphics[width=0.5\textwidth]{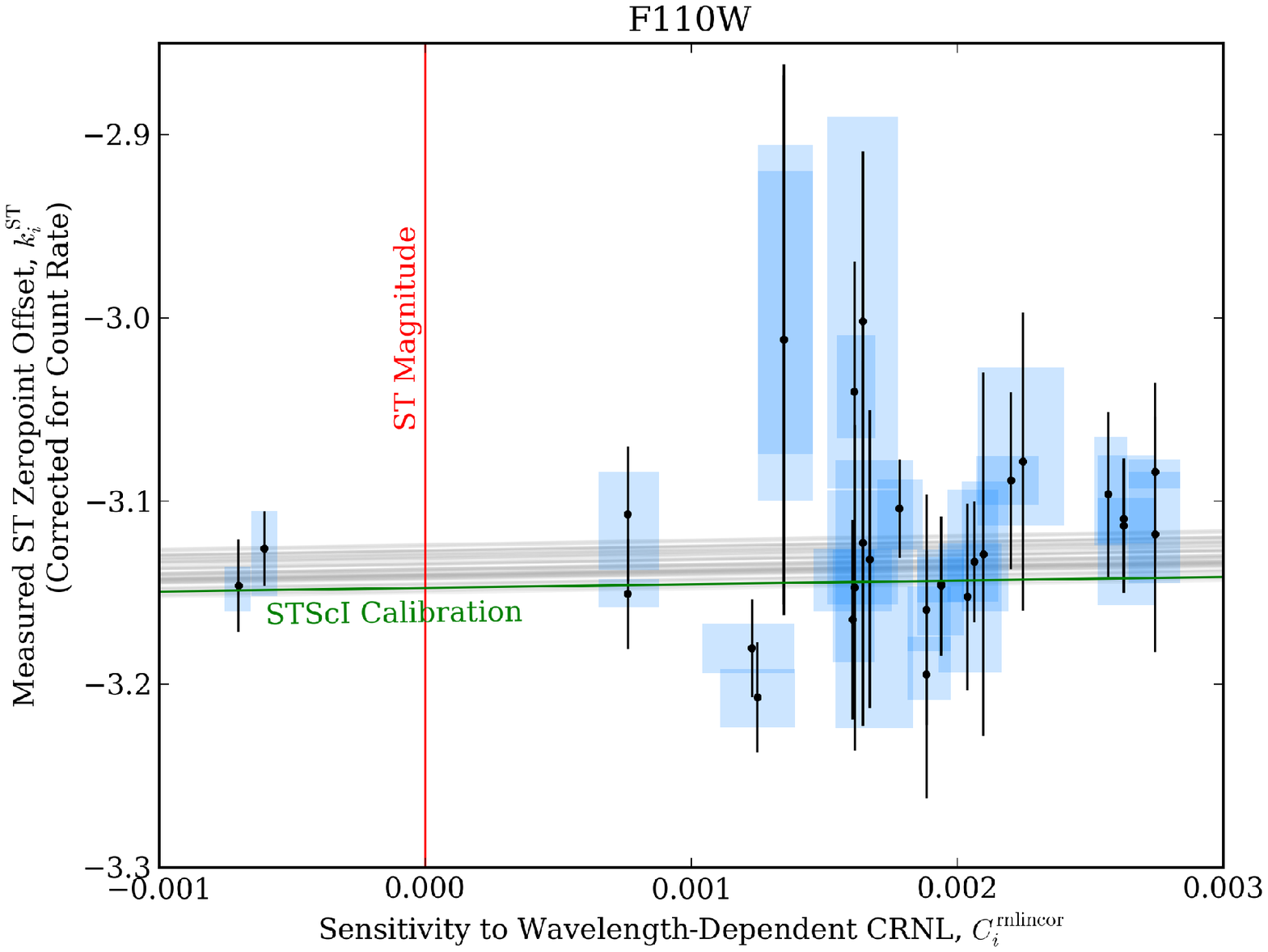}

\includegraphics[width=0.5\textwidth]{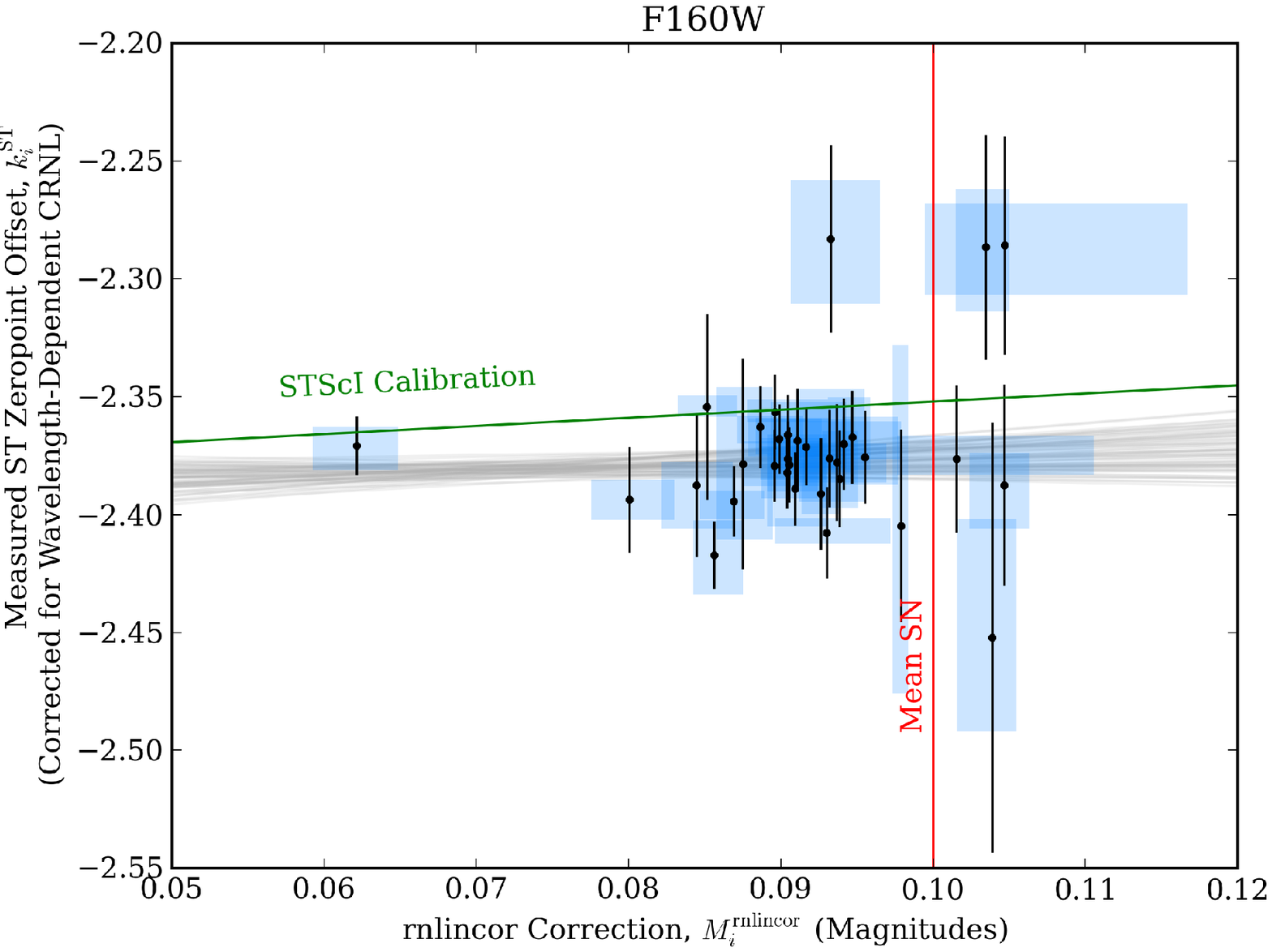}
\includegraphics[width=0.5\textwidth]{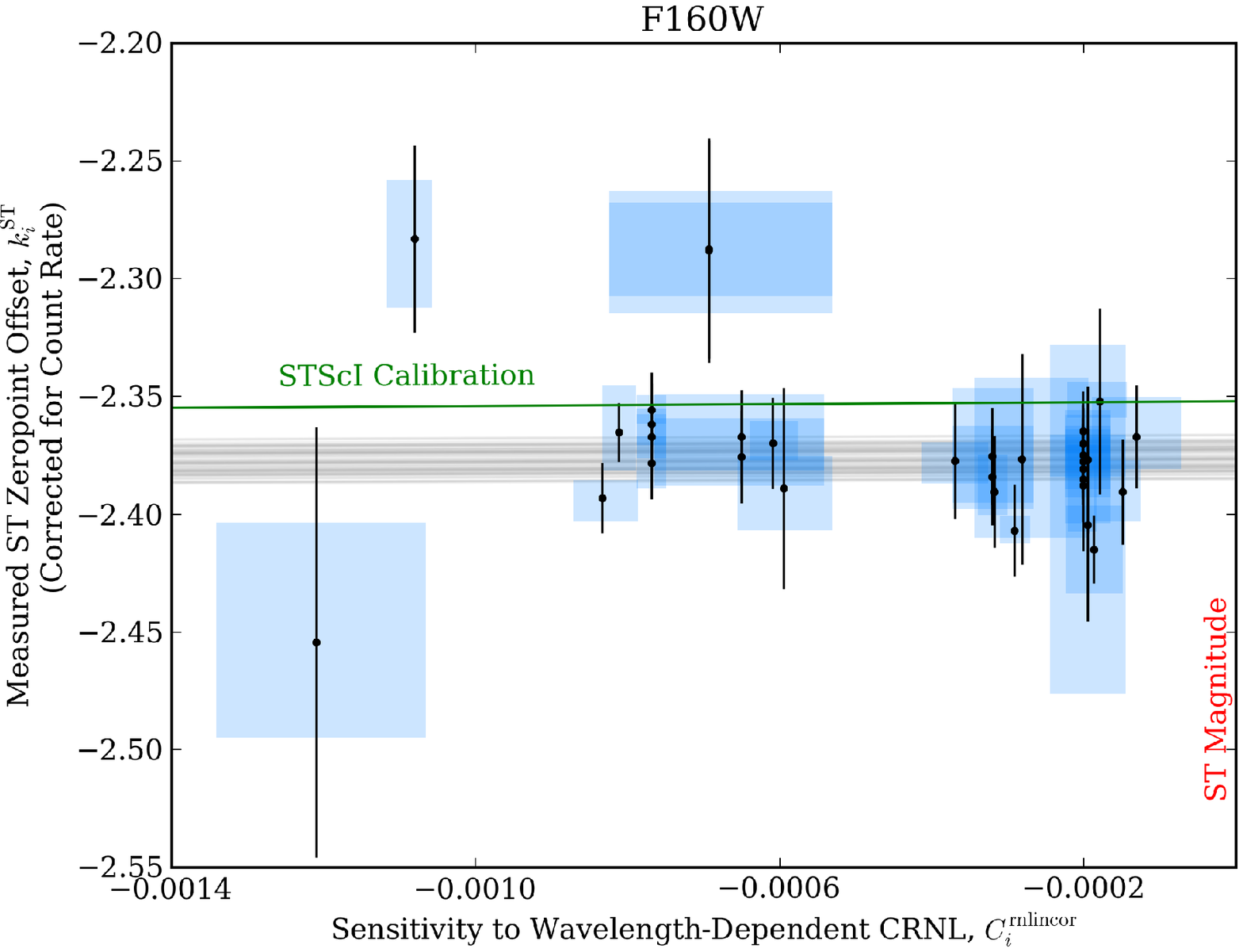}

\caption{
The top panels present the galaxy measurements for F110W; the bottom panels show F160W. The $y$-axis is always $k_i^{\mathrm{ST}}$, the ST zeropoint difference observed between NIC2 and WFC3, computed using \citet{brown14} templates. The left panels show $k_i^{\mathrm{ST}}$ plotted against the size (in magnitudes) of the \texttt{rnlincor} NIC2 correction for the galaxy; this is a measure of relative galaxy surface brightness, with higher surface brightness galaxies towards the left. The right panels show $k_i^{\mathrm{ST}}$ plotted against $C_i^{\mathrm{rnlincor}}$, the effect of the wavelength dependence of the CRNL on $k_i^{\mathrm{ST}}$ as the count-rate changes by one dex. This is a measure of relative galaxy color, with redder galaxies to the right. The blue boxes (one for each galaxy) represent the range in results for different analyses (e.g., varying the outer photometry radius) for each point. The black points and error bars represent the mean for that galaxy, with the mean error bar including $\sigma_{\mathrm{int}}$. Each gray line is the fit for each variant (Section~\ref{sec:uncertaintyanalysis}). The green lines present the STScI NIC2 and WFC3 calibrations, with the WFC3 IR zeropoints moved 0.04 magnitudes brighter (smaller) to represent the uncorrected WFC3 CRNL. In the left plots, the green lines are not horizontal as the WFC3 CRNL has not been corrected, and thus the expected NIC2/WFC3 zeropoint difference changes with count rate. In the right plots, the green lines are not horizontal as they show the ($\beta=2$) NIC2 bandpass shift to the red representing the preferential loss of blue sensitivity due to the NIC2 CRNL (Section~\ref{sec:nicmoseffectivebandpass}).
\label{fig:kST}}
\end{figure*}

\clearpage

\begin{deluxetable}{lrrrrrr}
\tablecolumns{7}
\tabletypesize{\scriptsize}
\setlength{\tabcolsep}{0.02in}
\tablecaption{Derived Galaxy Quantities\label{tab:calib}}
\tablehead{
 \colhead{Galaxy} &
 \colhead{NICMOS CRNL} &
 \colhead{Abscissa Color} &
 \colhead{Abscissa Value} &
 \colhead{$k$\tablenotemark{a}} &
 \colhead{$k_{\mathrm{ST}}^{\beta = 0}$ \tablenotemark{b}} &
 \colhead{$C_i^{\mathrm{rnlincor}}$}
 }
\startdata
\cutinhead{F110W, Post-NCS}

F110W\_01 & 0.232 &  F775W $-$ F110W & 1.273 & $-3.205 \pm 0.060$ & $-3.158 \pm 0.061$ & 0.00188 \\
F110W\_02 & 0.225 &  F775W $-$ F110W & 1.303 & $-3.196 \pm 0.031$ & $-3.148 \pm 0.032$ & 0.00194 \\
F110W\_03 & 0.225 &  F775W $-$ F110W & 1.355 & $-3.204 \pm 0.047$ & $-3.154 \pm 0.050$ & 0.00204 \\
F110W\_04 & 0.230 &  F775W $-$ F110W & 1.144 & $-3.082 \pm 0.068$ & $-3.040 \pm 0.069$ & 0.00161 \\
F110W\_05 & 0.233 &  F775W $-$ F110W & 1.171 & $-3.173 \pm 0.079$ & $-3.130 \pm 0.089$ & 0.00167 \\
F110W\_06 & 0.241 &  F775W $-$ F110W & 1.160 & $-3.160 \pm 0.098$ & $-3.118 \pm 0.099$ & 0.00165 \\
F110W\_07 & 0.267 &  F775W $-$ F110W & 1.016 & $-3.031 \pm 0.143$ & $-2.995 \pm 0.147$ & 0.00135 \\
F110W\_08 & 0.234 &  F775W $-$ F110W & 1.448 & $-3.140 \pm 0.044$ & $-3.087 \pm 0.044$ & 0.00220 \\
F110W\_09 & 0.220 &  F775W $-$ F110W & 1.369 & $-3.188 \pm 0.026$ & $-3.138 \pm 0.031$ & 0.00206 \\
F110W\_10 & 0.237 &  F775W $-$ F110W & 1.140 & $-3.203 \pm 0.051$ & $-3.161 \pm 0.053$ & 0.00161 \\
F110W\_11 & 0.253 &  F775W $-$ F110W & 1.145 & $-3.179 \pm 0.087$ & $-3.137 \pm 0.087$ & 0.00161 \\
F110W\_12 & 0.180 &  F814W $-$ F110W & -0.098 & $-3.138 \pm 0.005$ & $-3.149 \pm 0.007$ & -0.00060 \\
F110W\_13 & 0.211 &  F814W $-$ F110W & -0.125 & $-3.141 \pm 0.016$ & $-3.155 \pm 0.016$ & -0.00070 \\
F110W\_14 & 0.215 &  F775W $-$ F110W & 1.070 & $-3.218 \pm 0.018$ & $-3.187 \pm 0.019$ & 0.00123 \\
F110W\_15 & 0.218 &  F775W $-$ F110W & 0.766 & $-3.139 \pm 0.031$ & $-3.113 \pm 0.034$ & 0.00076 \\
F110W\_16 & 0.229 &  F775W $-$ F110W & 0.966 & $-3.242 \pm 0.023$ & $-3.207 \pm 0.025$ & 0.00125 \\
F110W\_17 & 0.213 &  F775W $-$ F110W & 1.240 & $-3.157 \pm 0.018$ & $-3.112 \pm 0.021$ & 0.00178 \\
F110W\_18 & 0.242 &  F775W $-$ F110W & 1.464 & $-3.169 \pm 0.061$ & $-3.112 \pm 0.062$ & 0.00274 \\
F110W\_19 & 0.233 &  F775W $-$ F110W & 1.410 & $-3.167 \pm 0.031$ & $-3.112 \pm 0.037$ & 0.00263 \\
F110W\_20 & 0.253 &  F775W $-$ F110W & 1.235 & $-3.117 \pm 0.079$ & $-3.068 \pm 0.081$ & 0.00225 \\
F110W\_21 & 0.264 &  F775W $-$ F110W & 1.172 & $-3.160 \pm 0.097$ & $-3.113 \pm 0.098$ & 0.00210 \\
F110W\_22 & 0.226 &  F775W $-$ F110W & 1.379 & $-3.152 \pm 0.040$ & $-3.098 \pm 0.043$ & 0.00257 \\

\cutinhead{F110W, Pre-NCS}

F110W\_61K\_01 & 0.169 &  F775W $-$ F110W & -0.065 & $-3.599 \pm 0.006$ & $-3.609 \pm 0.010$ & -0.00045 \\
F110W\_61K\_02 & 0.193 &  F775W $-$ F110W & -0.125 & $-3.567 \pm 0.013$ & $-3.582 \pm 0.016$ & -0.00067 \\
F110W\_61K\_03 & 0.191 &  F775W $-$ F110W & -0.076 & $-3.599 \pm 0.013$ & $-3.610 \pm 0.019$ & -0.00050 \\

\cutinhead{F160W, Post-NCS}

F160W\_01 & 0.091 &  F125W $-$ F160W & -0.054 & $-2.400 \pm 0.019$ & $-2.369 \pm 0.020$ & -0.00013 \\
F160W\_02 & 0.094 &  F125W $-$ F160W & -0.009 & $-2.407 \pm 0.016$ & $-2.371 \pm 0.016$ & -0.00061 \\
F160W\_03 & 0.105 &  F125W $-$ F160W & -0.309 & $-2.435 \pm 0.041$ & $-2.388 \pm 0.042$ & -0.00059 \\
F160W\_04 & 0.096 &  F125W $-$ F160W & -0.067 & $-2.422 \pm 0.016$ & $-2.376 \pm 0.018$ & -0.00065 \\
F160W\_05 & 0.105 &  F125W $-$ F160W & -0.068 & $-2.335 \pm 0.045$ & $-2.287 \pm 0.046$ & -0.00069 \\
F160W\_06 & 0.094 &  F125W $-$ F160W & -0.079 & $-2.414 \pm 0.017$ & $-2.385 \pm 0.019$ & -0.00032 \\
F160W\_07 & 0.104 &  F125W $-$ F160W & -0.153 & $-2.533 \pm 0.091$ & $-2.454 \pm 0.093$ & -0.00121 \\
F160W\_08 & 0.102 &  F125W $-$ F160W & -0.054 & $-2.413 \pm 0.029$ & $-2.377 \pm 0.029$ & -0.00019 \\
F160W\_09 & 0.094 &  F125W $-$ F160W & -0.122 & $-2.410 \pm 0.022$ & $-2.378 \pm 0.022$ & -0.00037 \\
F160W\_10 & 0.090 &  F125W $-$ F160W & -0.098 & $-2.423 \pm 0.009$ & $-2.369 \pm 0.011$ & -0.00077 \\
F160W\_11 & 0.086 &  F125W $-$ F160W & -0.083 & $-2.432 \pm 0.009$ & $-2.417 \pm 0.011$ & -0.00019 \\
F160W\_12 & 0.093 &  F125W $-$ F160W & -0.137 & $-2.430 \pm 0.016$ & $-2.408 \pm 0.016$ & -0.00029 \\
F160W\_13 & 0.087 &  F125W $-$ F160W & -0.198 & $-2.462 \pm 0.010$ & $-2.395 \pm 0.010$ & -0.00083 \\
F160W\_14 & 0.090 &  F125W $-$ F160W & -0.101 & $-2.379 \pm 0.013$ & $-2.366 \pm 0.013$ & -0.00020 \\
F160W\_15 & 0.093 &  F125W $-$ F160W & -0.308 & $-2.362 \pm 0.038$ & $-2.284 \pm 0.040$ & -0.00108 \\
F160W\_16 & 0.093 &  F125W $-$ F160W & -0.151 & $-2.413 \pm 0.021$ & $-2.392 \pm 0.022$ & -0.00032 \\
F160W\_17 & 0.062 &  F814W $-$ F160W & -0.292 & $-2.436 \pm 0.005$ & $-2.372 \pm 0.007$ & -0.00081 \\
F160W\_18 & 0.084 &  F125W $-$ F160W & -0.082 & $-2.399 \pm 0.028$ & $-2.388 \pm 0.029$ & -0.00020 \\
F160W\_19 & 0.080 &  F125W $-$ F160W & -0.063 & $-2.402 \pm 0.019$ & $-2.394 \pm 0.020$ & -0.00015 \\
F160W\_20 & 0.085 &  F125W $-$ F160W & -0.079 & $-2.365 \pm 0.038$ & $-2.355 \pm 0.038$ & -0.00018 \\
F160W\_21 & 0.088 &  F125W $-$ F160W & -0.128 & $-2.397 \pm 0.043$ & $-2.379 \pm 0.046$ & -0.00028 \\

\cutinhead{F160W, Pre-NCS}

F160W\_61K\_01 & 0.082 &  F125W $-$ F160W & -0.082 & $-2.677 \pm 0.044$ & $-2.666 \pm 0.045$ & -0.00020 \\
F160W\_61K\_02 & 0.078 &  F125W $-$ F160W & -0.063 & $-2.691 \pm 0.029$ & $-2.683 \pm 0.030$ & -0.00015 \\
F160W\_61K\_03 & 0.084 &  F125W $-$ F160W & -0.079 & $-2.666 \pm 0.060$ & $-2.656 \pm 0.061$ & -0.00018 \\
F160W\_61K\_04 & 0.088 &  F125W $-$ F160W & -0.128 & $-2.758 \pm 0.074$ & $-2.740 \pm 0.078$ & -0.00028 \\
F160W\_61K\_05 & 0.066 &  F110W $-$ F160W & -0.167 & $-2.758 \pm 0.004$ & $-2.710 \pm 0.006$ & -0.00065 \\
F160W\_61K\_06 & 0.078 &  F110W $-$ F160W & -0.197 & $-2.727 \pm 0.008$ & $-2.677 \pm 0.009$ & -0.00069 \\
F160W\_61K\_07 & 0.077 &  F110W $-$ F160W & -0.172 & $-2.752 \pm 0.007$ & $-2.704 \pm 0.011$ & -0.00067 \\

\enddata
\tablenotetext{a}{Instrumental magnitude difference between NICMOS and WFC3. The uncertainty is the mean statistical uncertainty on these measurements.}
\tablenotetext{b}{ST magnitude offset, computed using \citet{brown14} galaxy templates only. This uncertainty also includes variation due to photometry parameters.}

\end{deluxetable}

\section{Details of the Uncertainty Analysis}\label{sec:uncertaintyanalysis}

\subsection{Statistical Uncertainty}

The statistical uncertainties in the fits of $k_0^{\mathrm{ST}}$ (Equation~\ref{eq:meancalibration}) are \ErrorJStat in F110W and \ErrorHStat in F160W (both post-NCS). As the likelihood is approximately Gaussian, these are computed using the Jacobian matrices with the covariance matrix of observations.

\subsection{PSF Uncertainty}

Our PSFs, derived from P330E, are not identical to the PSFs of the galaxies. This will lead to systematic mismatches between the photometry for different filters. We verify our PSFs by varying the inner radius used (either 1 or 3 pixels / 0\farcs05 or 0\farcs15). We also vary the outer radius used (10 or 15 pixels/ 0\farcs5 or 0\farcs75). The range spanned by these changes is \ErrorJPSF in F110W and \ErrorHColCol in F160W, which we take as a systematic uncertainty. We also try a fully empirical PSF (not relying on Tiny Tim as a first approximation). This makes a difference of only 1 mmag.

\subsection{Impact of Other Zeropoints on the Color-Color Relations}
\label{sec:systOtherCal}

The slopes of the color-color relations used to calibrate F110W are $\sim 0.03$ mag/mag (with modest variation for different redshifts, templates, and abscissa colors), see Figure~\ref{fig:colorcolor}. This implies that the $\sim 0.03$ mag uncertainties on the ACS/WFC3 relative calibration (including zeropoints, encircled-energy correction, and the WFC3 CRNL) will contribute \ErrorJColCol to the uncertainty on the F110W calibration. For F160W, the slopes are $\sim 0.15$ mag/mag (again with modest variation), but the relative calibration uncertainties are smaller, as the abscissa colors generally both come from WFC3. We take a \ErrorHColCol uncertainty for this relation.

\subsection{Encircled Energy Correction}

Our measurement is sensitive to the differential in encircled energy between NIC2 and WFC3. Future updates to the encircled energy corrections can be propagated into our results; for the moment, we take a \ErrorJEE uncertainty.

\subsection{Annuli Correlations} \label{sec:anncorrelations}

The $C$ matrices that we empirically determine (Section~\ref{sec:instrumentalcolor}) have large off-diagonal correlations. These correlations are determined using object-free regions, and thus lack (smaller-scale) variations such as those caused by focus changes or sub-pixel position variations of sharp cores. As an approximate way to investigate the sensitivity to the ratio of small-scale to large-scale correlations in the $C$ matrices, we tried uniformly rescaling all of the off-diagonal elements by a range of values. These rescalings lower the dispersion in $k$ by more than a factor of two for stellar observations (these point-source observations show the largest response). There is little variation in the fitted zeropoint values or their error bars for a broad range of scale values, from 0.97 to 0 (where 0 results in an uncorrelated matrix). These rescalings have an effect of a few mmags on the post-NCS results (summarized in Table~\ref{tab:uncertainties}), which we take as systematic uncertainty.

\subsection{Uncertainty in Galaxy SEDs}
For the F160W bandpasses, the RMS residual from the color-color relation is \RMSHColCol, half of which we take as systematic uncertainty (to account for the fact that the average of our galaxies may not be the same as the average of the templates). Due to the similarities between the NIC2/WFC3 F110W bandpasses, the scatter in the color-color calibration for F110W is smaller (\RMSJColCol), as shown in Figure~\ref{fig:colorcolor}. We again take half this as systematic uncertainty. Switching to the \citet{bruzual03} templates changes the zeropoints by \ErrorJTem and \ErrorHTem (F110W and F160W post-NCS, respectively). It is also possible that our galaxies have more (or less) dust than the nearby galaxies used in constructing the \citet{brown14} templates. Adding 0.1 magnitude of CCM reddening \citep{cardelli89} to the templates (with $R_V = 3.1$, so $A_V = 0.31$) changes the zeropoints by \ErrorJEBV and \ErrorHEBV (F110W and F160W post-NCS, respectively). There is also uncertainty on the Milky Way foreground extinction for each galaxy, but these uncertainties affect our results at a trivial level.

\subsection{AGN Variability}

It is possible that some of our calibration galaxies have AGN, allowing them to change brightness in the time span between the NICMOS, ACS, and WFC3 observations. However, any large variability would flag the galaxy as unstable as we change the inner aperture size. Small variability is possible, but would increase $\sigma_{\mathrm{int}}$ in Equation~\ref{eq:meancalibration} and so is already included in the statistical error bar.

\subsection{WFC3 Uncertainties} \label{sec:WFCuncertainties}

Finally, we list WFC3 calibration uncertainties. As noted in Table~\ref{tab:highcountzeropoints}, we find 0.01 magnitudes of tension with the STScI zeropoints. Our zeropoints also scatter by a few mmags depending on aperture radius, and show some tension between standard stars. Until these issues are resolved, we take a 0.01 uncertainty in the WFC3 bright zeropoints. Going from bright to faint zeropoints adds about 0.01 magnitudes of uncertainty for the WFC3 CRNL \citep{riess10a, riess10b, riess11}, and moves the effective zeropoints 0.04 magnitudes brighter (lower). To be conservative, we also take half of our proposed update of the WFC3 bandpasses (Section~\ref{sec:wfceffectivebandpass} as uncertainty, giving 9~mmags in F110W and 7~mmags in F160W. In total, we estimate that the WFC3 low-count-rate ST zeropoints are $28.428\pm0.017$ for F110W and $28.176\pm0.016$ for F160W. Note that these zeropoints are tied to the CALSPEC system, which has uncertainties as well.

\bibliographystyle{apj}
\bibliography{ms}

\end{document}